\documentclass[twocolumn,secnumarabic,amssymb,nobibnotes,aps,showpacs,nofootinbib,noeprint]{revtex4-1}
\usepackage[english]{babel}
\usepackage{graphicx}

\usepackage{amsmath}
\usepackage{amsfonts}
\usepackage[usenames,dvipsnames]{color}
\usepackage{hyperref}
\usepackage{rotating}
\usepackage{slashed}
\usepackage{transparent}

\DeclareMathAlphabet{\mathbb}{U}{bbold}{m}{n}

\begin{document}
\title{Anisotropic pair correlations in binary and multicomponent hard-sphere mixtures in the vicinity of a hard wall: 
A combined density functional theory and simulation study}
\author{Andreas H\"{a}rtel,$^{1,*}$ Matthias Kohl,$^{2,*}$ and Michael Schmiedeberg$^2$} 
\affiliation{$^1$ Institut of Physics, 
Johannes Gutenberg-University Mainz, 
Staudinger Weg 9, 55128 Mainz, Germany\\
$^2$ Institute for Theoretical Physics II: Soft Matter, 
Heinrich-Heine University D\"usseldorf, Universit\"atsstr. 1, 40225 D\"usseldorf, Germany}
\pacs{82.70.Dd,61.20.-p,64.70.qd}
\date{\today}

\footnotetext[1]{A.H. and M.K. contributed equally to this work. 
A.H. dealt with the DFT calculations; M.K., with the simulations. All authors wrote the paper. }

\begin{abstract}
The fundamental measure approach to classical density functional theory has been shown to be a 
powerful tool to predict various thermodynamic properties of hard-sphere systems. We employ 
this approach to determine not only one-particle densities but also two-particle correlations 
in binary and six-component mixtures of hard spheres in the vicinity of a hard wall. The broken 
isotropy enables us to carefully test a large variety of theoretically predicted two-particle 
features by quantitatively comparing them to the results of Brownian dynamics simulations. 
Specifically, we determine and compare the one-particle density, the total correlation functions, 
their contact values, and the force distributions acting on a particle. 
For this purpose, we follow the compressibility route and theoretically calculate the direct 
correlation functions by taking functional derivatives. We usually 
observe an excellent agreement between theory and simulations, except for small deviations in 
cases where local crystal-like order sets in. Our results set the course for further 
investigations on the consistency of functionals as well as for structural analysis on, e.g., 
the primitive model. In addition, we demonstrate that due to the 
suppression of local crystallization, the predictions of six-component mixtures are better 
than those in bidisperse or monodisperse systems. Finally, we are confident 
that our results of the structural modulations induced by the wall lead to a deeper understanding 
of ordering in anisotropic systems in general, the onset of heterogeneous crystallization, 
caging effects and glassy dynamics close to a wall, as well as structural properties in systems 
with confinement.
\end{abstract}

\maketitle

%%%%%%%%%%%%%%%%%%%%%%%%%%%%%%%
%SECTION 1 %%%%%%%%%%%%%%%%%%%%
%%%%%%%%%%%%%%%%%%%%%%%%%%%%%%%
\section{Introduction}

In order to study the structure or dynamics of simple fluids or liquids, usually model 
systems consisting of particles that interact according to simple pair potentials are 
considered. A large variety of phenomena can be explored in such model systems, e.g., 
interfaces between different fluid phases \cite{tarazona_mph54_1985} or between a liquid and a 
vapor \cite{Wertheim1976,evans_molphys80_1993,evans_jpcm21_2009,parry_jpcm26_2014} as well 
as phase transitions between fluids and solids \cite{oettel_pre82_2010,haertel_prl108_2012,haertel_book_2013}. Furthermore, glassy dynamics or jamming effects can be observed for such systems 
at large packing fractions or low temperatures \cite{Liu1998,liu10,hunter12,lu13}. 
The relation of the slowdown of dynamics and structural properties is the subject of 
ongoing research (see, e.g., 
\cite{Pusey1986,Scheidler2002,Mayer2008,Mittal2008,Goel2009,royall_jncs407_2014,Royall2015a,dunleavy_nc6_2015}).

One of the most important particulate model systems is the simple hard-sphere (HS) 
system, where overlaps of two particles are not allowed and spheres do not directly 
interact if they do not overlap. HS systems not only serve as a simple model system, but 
also are used as reference systems. For example, the structure of simple fluids 
with more complex interactions often is compared to the structure of HSs with an effective 
diameter \cite{rowlinson64,barker67,Andersen1971}. Furthermore, the dynamics 
of spheres with purely repulsive, finite-ranged interaction can be mapped onto the dynamics 
of HSs \cite{Xu2009,Schmiedeberg2011,Haxton2011}. 

In this paper we investigate an HS system in the vicinity of a hard wall. We usually 
consider a bidisperse system that does not crystallize, but also present some results for 
monodisperse and six-component dispersions. Since we are especially interested in the 
anisotropic order induced by the wall, we not only study the one-particle density, but 
also determine the two-particle correlation functions. The structural modulations and local 
ordering in the vicinity of a wall is of great interest in order to understand the onset 
of heterogeneous crystallization \cite{winter09,sandomirski14}
and in order to obtain deeper insight into the influence of local order on the complex 
dynamics close to a wall \cite{heni00,Scheidler2002,Scheidler2004,allahyarov_natcom6_2015} or even in confinement \cite{kurzidim09,Lomba2014}. 
In this paper we also use the broken symmetry of the system in order to quantitatively 
test the two-particle predictions of the fundamental measure theory (FMT) approach to classical 
density functional theory (DFT) via the compressibility route as explained in the following. 

DFT was originally developed by Hohenberg and Kohn for an electron gas at zero 
temperature \cite{hohenberg_prv136_1964} and later extended for nonzero 
temperatures \cite{mermin_prv137_1965}. In the meantime, DFT for classical systems has been formulated 
and it has turned out to be a powerful tool in order to predict thermodynamic properties 
of classical systems, especially in the field of soft matter, e.g., for fluid many-body 
systems \cite{ebner_pra14_1976,evans_ap28_1979,wu_aiche52_2006}. DFT even was employed to study 
crystallization \cite{baus_jpcm2_1990,tarazona_prl84_2000,roth_jpcm14_2002,hansen-goos_jpcm18_2006,oettel_pre82_2010,neuhaus13c,neuhaus14}, 
interfaces between a crystal and a fluid \cite{haertel_prl108_2012,oettel_pre82_2010,oettel_pre86_2012} 
and complex ordering of particles due to interactions with multiple length scales \cite{Mladek2006,Likos2007,archer13,archer15}
or external potentials \cite{haertel_pre81_2010,neuhaus13,neuhaus13b}, as well as to explore dynamical phenomena \cite{marconi99,archer04,espanol09,haertel_pre81_2010,neuhaus13b,Lichtner14}. 

A fundamental approach in order to obtain a suitable free energy functional for an HS 
system was introduced by Rosenfeld with the so-called FMT \cite{rosenfeld_prl63_1989}. 
Different versions of FMT have been presented in the meantime \cite{tarazona_prl84_2000,roth_jpcm14_2002,yu_jcp117_2002,hansen-goos_jpcm18_2006,santos12,hansen-goos_pre91_2015}, 
including the functional known as the White Beak mark II (WBII) 
functional \cite{hansen-goos_jpcm18_2006}, 
which has been extensively employed and tested in order to predict one-particle 
densities \cite{hansen-goos_jpcm18_2006,oettel_pre82_2010,lang_prl105_2010,haertel_prl108_2012,neuhaus14}. 
Two-particle correlations are attainable via the test-particle and 
the compressibility route, which lead to consistent results in the case of the exact (but unknown) 
free energy functional. Here we use the compressibility route and the WBII functional in order 
to calculate two-particle correlations in a 
system that is not isotropic due to the proximity of a wall. The theoretical predictions 
are compared to the results obtained from Brownian dynamics (BD) simulations. We observe 
an excellent agreement as long as local crystal-like structures are avoided. 

The paper is organized as follows: In Sec.~\ref{sec:section2} the model system is introduced 
and explained. The simulation details are presented in Sec.~\ref{sec:section3}. 
In Sec.~\ref{sec:section4} we discuss how the one- and two-particle correlations are 
obtained within our FMT approach. The results are presented and compared to our 
simulation data in Sec.~\ref{sec:section5}.
Finally, we conclude in Sec.~\ref{sec:section6}

%%%%%%%%%%%%%%%%%%%%%%%%%%%%%%%
%SECTION 2 %%%%%%%%%%%%%%%%%%%%
%%%%%%%%%%%%%%%%%%%%%%%%%%%%%%%
\section{Model System: Spheres close to a wall}
\label{sec:section2}

\begin{figure}
\includegraphics[width=8.0cm]{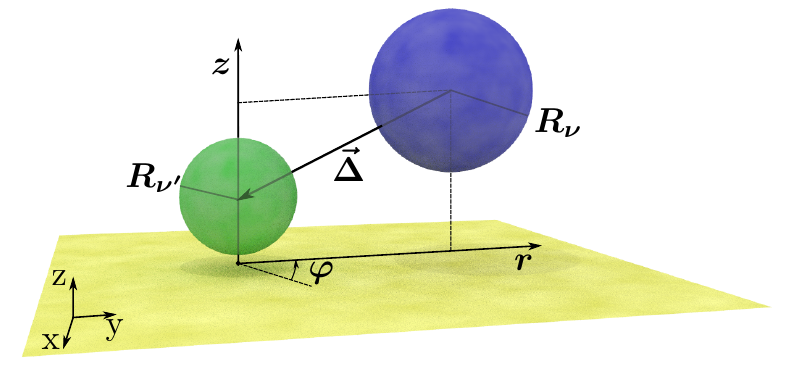}
\caption{\label{fig:Figure1}(Color online) 
Sketch of two hard spheres of different species, $\nu$ and $\nu'$, close to a hard wall. 
Their respective positions $\vec{r}$ and $\vec{r}{\,}'$ (not shown) define their relative 
distance $\vec{\Delta}=\vec{r}{\,}'-\vec{r}$. Particle diameters are $\sigma_{\nu}=2R_{\nu}$ 
and $\sigma_{\nu'}=2R_{\nu'}$. We employ cylindrical coordinates $z$, $r$, $\varphi$ 
around the left (green) sphere. }
\end{figure}

We consider multicomponent mixtures of HS suspended in a homogeneous 
solvent next to a flat hard wall. The solvent is integrated out and only contributes to 
the stochastic overdamped Brownian motion of the colloidal particles. We investigate 
monodisperse, binary, and six-component mixtures in equilibrium, which we access with both 
classical DFT and BD computer simulations. 
For the latter, the number of particles of each species $\nu$ is fixed, i.e., at a 50:50 mixture in the case of a binary system. In the grand canonical framework of DFT all species are 
assumed to have the same averaged number densities in a reference bulk system. In the case of the binary system, the spheres have diameters $\sigma_1$ and $\sigma_2=1.4\sigma_1$ in order to avoid crystallization effects \cite{OHern2003}.

The wall is located in the $xy$ plane at position $z=0$ (see Fig.~\ref{fig:Figure1}). 
To express two-particle correlations, we consider one sphere at position $(x',y',z')$ as the reference particle such that the positions $(x,y,z)$ of all other particles can be expressed in cylindrical coordinates relative to the reference sphere. As a consequence, two-particle correlations depend on the distance $z'$ of the reference sphere to the wall, the distances $z$ of the other particles to the wall, and the distance $r=\left[(x-x')^2+(y-y')^2\right]^{1/2}$ between the particles and the reference sphere measured parallel to the wall. All other coordinates are integrated out due to symmetry. As a consequence, no crystallization or other symmetry-breaking ordering parallel to the wall is resolved.

We compare one- and two-particle statistical averages. For example, the one- and two-particle densities are defined as 
\begin{align} 
\rho_{\nu}(\vec{r}) &=  
	\left\langle \sum\limits_i \delta(\vec r-\vec r_{\nu,i})\right\rangle, 
\label{eq:one-particle-density} \\
\rho_{\nu\nu'}^{(2)}(\vec{r},\vec{r}{\,'}) 
&= \left\langle {\sum_{i,i'}}' \delta\left(\vec{r}-\vec{r}_{\nu,i}\right)
\delta\left(\vec{r}{\,'}-\vec{r}_{\nu',i'}\right) \right\rangle , 
\label{eq:two-particle-density}
\end{align}
where $\langle \ldots \rangle$ denotes the ensemble average (canonical in simulations and grand canonical in DFT). The primed sum $\sum_{i,i'}^{'}$ runs over all species $\nu$ and $\nu'$ and particles $i=1 \ldots N_\nu$ 
with $i\neq i'$ in the case of $\nu=\nu'$. The packing fraction is given by
$\phi=\sum_\nu \phi_\nu= \sum_\nu \tfrac{\pi}{6}\sigma_\nu^3\rho_\nu$.

%%%%%%%%%%%%%%%%%%%%%%%%%%%%%%%
%SECTION 3 %%%%%%%%%%%%%%%%%%%%
%%%%%%%%%%%%%%%%%%%%%%%%%%%%%%%
\section{Simulations}
\label{sec:section3}

\subsection{Brownian dynamics}

\begin{figure}
\centering
\includegraphics[width=8.0cm]{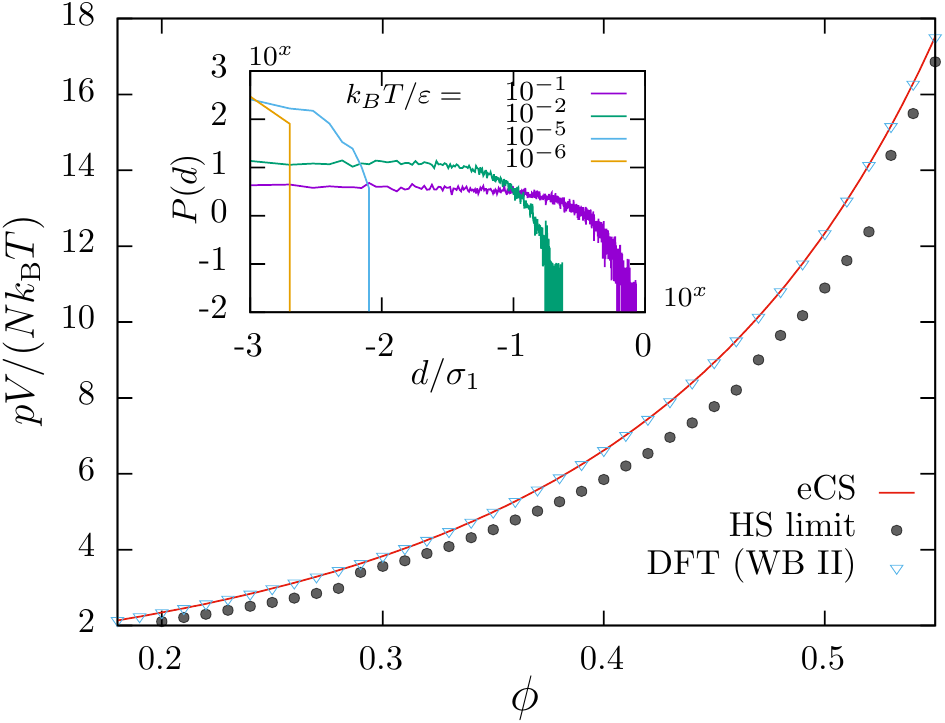}
\caption{\label{fig:Figure2}(Color online) 
Equation of state for a binary and equimolar 
hard-sphere (HS) system, i.e., $pV/(Nk_{{\rm B}}T)$ as a function of the total packing fraction $\phi$.
Triangles are based on calculations within DFT using the WBII approximation (see Sec. \ref{sec:section4}) and the solid (red) line is based on predictions from the extended Carnahan-Starling (eCS) equation \cite{Mansoori1971,Malijevsky1999}. Filled circles denote the results of Brownian dynamics simulation in the HS limit. Inset: The normalized probability distribution of finding two particles with an overlap 
$d$ for different temperatures in a double-logarithmic representation. }
\end{figure}

To test the theoretical calculations we employ BD simulations (see, e.g., \cite{Allen1999}) which are based
 on the overdamped Langevin equation, 
\begin{equation}
\gamma_\nu \dot{\vec r}_{\nu,i}(t) = {\vec f}_{\nu,i}\big(\left\{\vec r_{\nu',1},\ldots, \vec r_{\nu',N_{\nu'}}\right\}_{\nu'=1,2,...}\big) + \vec \xi_{\nu,i}(t)\, , \label{eq:langevin}
\end{equation}
where $\gamma_\nu$ is the friction constant that we consider to be proportional to the diameter $\sigma_{\nu}$ of the spheres. The force ${\vec f}_{\nu,i}$ includes all forces due to pair interactions and the external field. 
In addition, a random Gaussian force $\vec \xi_{\nu,i} (t)$ is acting 
on the particles. The first moment of the distribution of random forces is $0$, whereas the second moment fulfills the 
fluctuation dissipation relation, i.e.,
$\big\langle \vec \xi_{\nu,i}(t) \vec{\xi}_{\nu',i'}^\text{T}(t') \big\rangle = 2 \gamma_\nu k_{\rm B} T \delta_{\nu\nu'}\delta_{ii'}\delta(t-t')\tensor{{\rm I}}$, 
with $k_{\rm B}T$ being the product of the temperature $T$ and Boltzmann constant $k_{\rm B}$, $\vec{\xi}_{\nu'}^\text{T}$ being the transpose of $\vec{\xi}_{\nu'}$, and 
$\tensor{{\rm I}}$ the three-dimensional unit matrix. $\delta(t-t')$ and $\delta_{\nu\nu'}\delta_{ii'}$ 
stand for the Dirac $\delta$ distribution and two Kronecker $\delta$, respectively. 

We employ a cubic simulation box with side length $l$, periodic boundary conditions 
in the $x$ and $y$ direction, and walls at $z=0$ and $z=l$. We use $N=32000$ particles, such that the box is large enough to avoid confinement effects such as nontrivial correlations of particles with both walls.

\subsection{Hard-sphere limit}

Molecular dynamics \cite{Xu2009} and BD studies \cite{Lopez-Flores2013} 
have shown that with decreasing temperature all properties of a system with finite-ranged 
and purely repulsive interactions approaches well-defined limiting values that 
coincide with the properties of hard-sphere mixtures and therefore is called the HS limit in the following. 

In our simulations we apply the soft and purely repulsive pair potential 
\begin{equation}
 u_{\nu\nu'}(\Delta)= \left\{ \begin{array}{lcl}%
                             \frac{\varepsilon}{2}\left( 1- \frac{\Delta}{\sigma_{\nu\nu'}}\right)^2 & \quad & \Delta\leq \sigma_{\nu\nu'} \\%
                             0 & & \text{otherwise} ,
                            \end{array}%
\right.
 \label{eq:pair_potential}
\end{equation}
where $\sigma_{\nu\nu'}=(\sigma_\nu + \sigma_{\nu'})/2$ is the intermediate diameter and the prefactor $\varepsilon$ sets the energy scale. At sufficiently low temperatures, where $\varepsilon/k_\text{B}T\gg 1$, 
the particle overlaps become very small and particles interact like HSs. 
We consider our system to be in the HS limit if the average overlap of 
two interacting particles is smaller than $5\%$. The double-logarithmic inset in 
Fig.~\ref{fig:Figure2} shows how the probability distribution $P(d)$ of 
overlaps $d$ converges with decreasing temperature against a very narrow $\delta(d)$-like distribution. 
In the main plot in Fig.~\ref{fig:Figure2} we show the equation of state of a binary HS mixture, calculated by means of our DFT as well as from the predictions of 
Boubl\'ik \cite{Boublik1970} and  Mansoori \emph{et al.} \cite{Mansoori1971}; the latter 
is also known as the extended Carnahan-Starling equation of state \cite{Malijevsky1999}. 
We compare these curves with the measured virial pressure from our BD simulations in the HS limit. 
Due to the small remaining overlaps in our simulations, we usually obtain a very slightly deviating 
pressure in comparison to the theoretical predictions, while it is known that the 
structural \cite{rowlinson64,barker67,Andersen1971} as well as 
the dynamical \cite{Xu2009,Schmiedeberg2011,Haxton2011,Lopez-Flores2013} properties 
are even closer to the HS system. 
The formalism of the theoretical DFT calculations are presented in the next section.

%%%%%%%%%%%%%%%%%%%%%%%%%%%%%%%
%SECTION 4 %%%%%%%%%%%%%%%%%%%%
%%%%%%%%%%%%%%%%%%%%%%%%%%%%%%%
\section{Density Functional Theory}
\label{sec:section4}

In this section we discuss classical DFT within the framework of 
FMT \cite{rosenfeld_prl63_1989,roth_jpcm22_2010}
leading to direct particle 
correlations \cite{evans_ap28_1979,tarazona_inbook_2008,hansen_book_2013}. 
The Ornstein-Zernike (OZ) relation links them with the total correlations between particles. 
We introduce the theory for a multicomponent HS system in a geometry where isotropy is broken due to the wall.

\subsection{DFT for multi-component systems}

In the framework of (classical) DFT 
\cite{evans_ap28_1979,tarazona_inbook_2008}, 
a functional, $\Omega[\{\rho_\nu\}]\equiv\Omega(T,V,\{\mu_\nu\};[\{\rho_\nu\}])$, 
of the sets $\{\rho_{\nu}\}$ of one-particle densities $\rho_{\nu}$ and $\{\mu_\nu\}$ of chemical potentials $\mu_\nu$ for species $\nu=1 \ldots n$ 
can be defined at fixed external potential $V^{\rm ext}=\sum_{\nu} V_{\nu}^{\rm ext}$ 
such that the grand canonical potential $\Omega\equiv\Omega(T,V,\{\mu_\nu\})$ 
is obtained when the set 
$\{\rho_\nu^{\rm (eq)}\}$ of equilibrium one-particle densities 
is used as an input. The functional can be written as 
\begin{equation}
\Omega[\{\rho_\nu\}] = 
{\cal F}[\{\rho_\nu\}] - \sum_{\nu'=1}^{n} \int_V \rho_{\nu'}(\vec{r}{\,'}) \psi_{\nu'}(\vec{r}{\,'}) d\vec{r}{\,'} , 
\label{eq:gc_functional}
\end{equation}
where the intrinsic free energy functional ${\cal F}$ 
and the intrinsic chemical potentials 
$\psi_{\nu'}$ as unique functionals of the 
one-particle densities $\rho_{\nu}$ have been introduced. 

The grand canonic functional in Eq.~(\ref{eq:gc_functional}) 
has the property to be minimized by the equilibrium one-particle densities, 
thus, its functional derivative vanishes for each species $\nu'$, i.e., for all $\nu'$, 
\begin{equation}
\left.\frac{\delta\Omega[\{\rho_\nu\}]}{\delta \rho_{\nu'}(\vec{r}{\,'})}
\right|_{\big\{\rho_\nu\big\}=\big\{\rho_{\nu}^{\rm (eq)}\big\}} = 0 \, . 
\end{equation}
Accordingly, the intrinsic chemical potentials read 
\begin{equation}
\psi_{\nu'}\left( \vec{r}{\,'}; [\left\{\rho_\nu\right\}] \right)
= \frac{\delta {\cal F}[\{\rho_\nu\}]}{\delta\rho_{\nu'}(\vec{r}{\,'})} . 
\label{eq:intrinsic-chemical-potential}
\end{equation}
Furthermore, the free energy of the system is defined as the sum of the intrinsic free energy ${\cal F}$ and 
the energy due to the external potential, 
\begin{equation}
F = {\cal F}\left[\left\{\rho_{\nu}^{\rm (eq)}\right\}\right] 
+ \sum_{\nu'=1}^{n} \int_V \rho_{\nu'}^{\rm (eq)}(\vec{r}{\,'}) V_{\nu'}^{\rm ext}(\vec{r}{\,'}) d\vec{r}{\,'} . 
\end{equation}
On the other hand, the free energy also follows from the grand canonical potential via a Legendre 
transform, $\Omega=F-\sum_{\nu'}\mu_{\nu'} N_{\nu'}$. 
Together with Eq.~(\ref{eq:intrinsic-chemical-potential}), this leads to 
\begin{equation}
\mu_{\nu'} = V_{\nu'}^{\rm ext}(\vec{r}{\,'}) 
+ \psi_{\nu'}\left( \vec{r}{\,'}; \left[\left\{\rho_\nu^{\rm (eq)}\right\}\right] \right) , 
\label{eq:fundamental_equation}
\end{equation}
which in \cite{evans_ap28_1979} is termed ``the fundamental equation in the theory of non-uniform liquids''.
Together with the representations of the intrinsic chemical potentials in 
Eq.~(\ref{eq:intrinsic-chemical-potential}), one can use Eq. (\ref{eq:fundamental_equation}) as an implicit 
equation to determine the equilibrium densities $\rho_\nu^{\rm (eq)}$, if the intrinsic free 
energy functional ${\cal F}[\{\rho_\nu\}]$ 
is known. 

The case of the non-interacting particles of an ideal gas is well known: The intrinsic free energy is 
\begin{equation}
{\cal F}^{\rm id}[\{\rho_\nu\}] = k_{\rm B}T \sum_{\nu'=1}^{n} \int_V \rho_{\nu'}(\vec{r}{\,'}) 
\left[ \ln\left(\rho_{\nu'}(\vec{r}{\,'})\Lambda_{\nu'}^3\right)-1\right] d\vec{r}{\,'} , 
\label{eq:free_energy_ideal}
\end{equation}
which leads to the equilibrium density profiles 
$\rho_\nu^{\rm (eq)}(\vec{r})=z_\nu \exp(-\beta V_\nu^{\rm ext}(\vec{r}))$ 
with the fugacities $z_{\nu}=\exp(\beta\mu_\nu)\Lambda_\nu^{-3}$, 
the (irrelevant) thermal wavelengths $\Lambda_{\nu}$, and 
the inverse temperature $\beta=1/k_{\rm B}T$, containing the product of the temperature $T$ and Boltzmann constant $k_{\rm B}$. 

In the case of systems with interacting particles, it is common to split the intrinsic free energy functional 
\begin{equation}
{\cal F}[\{\rho_\nu\}] = {\cal F}^{\rm id}[\{\rho_\nu\}] + {\cal F}^{\rm exc}[\{\rho_\nu\}]
\label{eq:F_split_up}
\end{equation}
into the known ideal-gas part from Eq.~(\ref{eq:free_energy_ideal}) and an 
overideal excess part ${\cal F}^{\rm exc}$ which includes all particle interactions. 
Consequently, putting Eqs.~(\ref{eq:intrinsic-chemical-potential}), (\ref{eq:free_energy_ideal}), 
and (\ref{eq:F_split_up}) together, the dimensionless intrinsic chemical potential becomes 
\begin{equation}
 \beta \psi_{\nu'}\big(\vec{r}{\,'};[\{\rho_\nu\}]\big)=\ln(\rho_{\nu'}\Lambda_{\nu'}^3)-c_{\nu'}^{(1)}\big(\vec{r}{\,'};[\{\rho_\nu\}]\big),
 \label{eq:intrinsic_chem_potential}
\end{equation} 
where
\begin{align}
c_{\nu'}^{(1)}\big(\vec{r}{\,'};[\{\rho_\nu\}]\big) &= -\beta 
\frac{\delta {\cal F}^{\rm exc}[\{\rho_\nu\}]}
{\delta\rho_{\nu'}(\vec{r}{\,'})} , \label{eq:direct-correlations1}
\end{align}
is the first member of a hierarchy of direct correlation functions which contain 
full information on the structural properties of the corresponding system. The next member of the hierarchy reads 
\begin{align}
c_{\nu'\nu''}^{(2)}\big(\vec{r}{\,'},\vec{r}{\,''};[\{\rho_\nu\}]\big) &= -\beta 
\frac{\delta^2 {\cal F}^{\rm exc}[\{\rho_\nu\}]}
{\delta\rho_{\nu'}(\vec{r}{\,'})\delta\rho_{\nu''}(\vec{r}{\,''})} . 
\label{eq:direct-correlations2}
\end{align}

By inserting Eq.~(\ref{eq:intrinsic_chem_potential}) into Eq.~(\ref{eq:fundamental_equation}), 
we get a formal solution for the density profiles which reads 
\begin{equation}
\rho_{\nu'}^{\rm (eq)}(\vec{r}{\,'})
=z_{\nu'} \exp\left(-\beta V_{\nu'}^{\rm ext}(\vec{r}{\,'})+c_{\nu'}^{(1)}\big(\vec{r}{\,'};[\{\rho_\nu^{\rm (eq)}\}]\big)\right) . 
\label{eq:ele-density-profile}
\end{equation}
This equation provides an iterative procedure for minimizing the grand canonical functional: 
Starting from random initial density profiles, Eq.~(\ref{eq:ele-density-profile}) can be applied 
repeatedly in order to approach the equilibrium density profile numerically (Picard iteration).

\begin{figure}
\centering
\includegraphics[width=8.3cm]{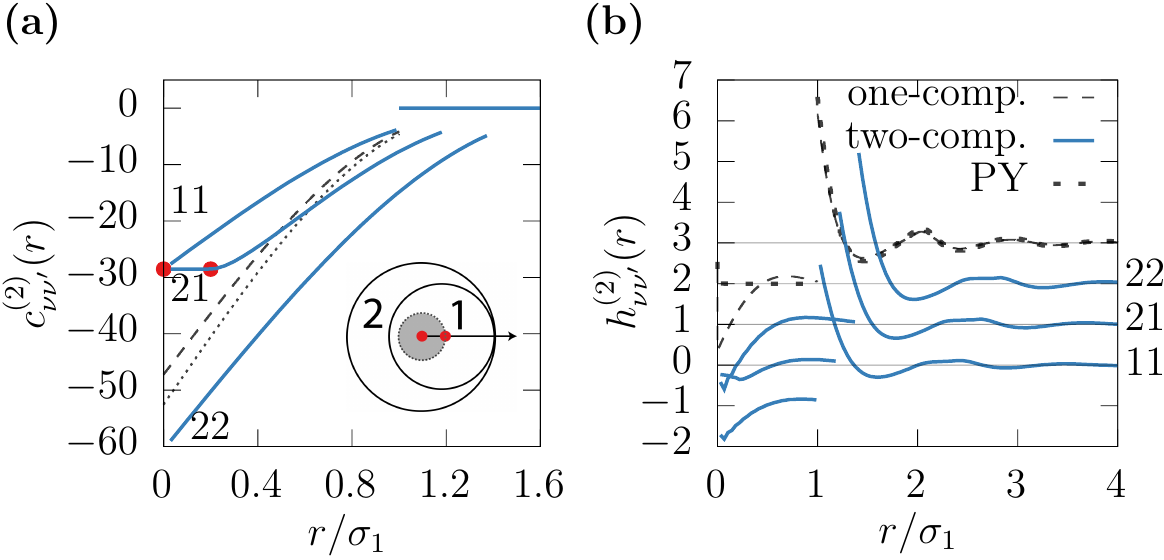}
\caption{\label{fig:Figure3}(Color online) 
(a) Direct and (b) total correlation functions in bulk for one- and two-component (50:50) 
hard-sphere systems with volume fraction $\phi=0.5$. In the two-component case, the correlations 
between possible combinations of species are labeled $11$ (small-small), 
$21$ (large-small), and $22$ (large-large). All correlations are determined from our 
DFT calculations in combination with the OZ relation; for comparison we also show the 
analytically known Percus-Yevick (PY) result for the one-component system \cite{wertheim_prl10_1963}. 
Inset in (a): Sketch showing that if a small particle 1 is inside a 
larger particle 2, its center point can move within the shaded (gray) area without changing the intersection 
volume of the spheres. As a consequence, there is a plateau in the 21 curve between the two filled (red) circles. }
\end{figure}

\subsection{Ornstein-Zernike relation}
\label{sec:OZ-relation}

The pair-distribution function is given via 
the one- and two-particle densities as defined
in Eqs.~(\ref{eq:one-particle-density}) and (\ref{eq:two-particle-density}) by (see, e.g., \cite{hansen_book_2013}) 
\begin{equation}
g_{\nu\nu'}^{(2)}(\vec{r},\vec{r}{\,'}) = 
\frac{\rho_{\nu\nu'}^{(2)}(\vec{r},\vec{r}{\,'})}{\rho_{\nu}(\vec{r})\rho_{\nu'}(\vec{r}{\,'})} . 
\label{eq:pair-distribution-function}
\end{equation}
The total correlation function $h$ is defined by 
\begin{equation}
h_{\nu\nu'}^{(2)}(\vec{r},\vec{r}{\,'}) = g_{\nu\nu'}^{(2)}(\vec{r},\vec{r}{\,'}) - 1 .
\end{equation}
It is related to the direct correlation function $c_{\nu\nu'}^{(2)}(\vec{r},\vec{r}{\,'})$ as defined in Eq. (\ref{eq:direct-correlations2}) via the OZ relation \cite{hansen_book_2013}, 
\begin{align}
h_{\nu\nu'}^{(2)}(\vec{r},\vec{r}{\,'}) &= c_{\nu\nu'}^{(2)}(\vec{r},\vec{r}{\,'}) 
\label{eq:ornstein-zernike-equation} \\
+& \sum_{\nu''=1}^{n} \int_V h_{\nu\nu''}^{(2)}(\vec{r},\vec{r}{\,''}) 
\rho_{\nu''}(\vec{r}{\,''}) c_{\nu''\nu'}^{(2)}(\vec{r}{\,''},\vec{r}{\,'}) d\vec{r}{\,''} \notag . 
\end{align}

Forestalling results from DFT calculations that are explained later, both kinds of correlation functions are illustrated in Fig.~\ref{fig:Figure3} for a monodisperse and a binary 
system in bulk. In the binary system, four combinations 
between small and large particles exist, where the mixed combinations small-large and large-small 
are identical in bulk. The direct correlations are calculated using FMT as described in the next subsection. We obtain the total correlations via the OZ relation by employing the direct correlations and their corresponding density profiles. 
This method via the direct correlations is called the compressibility route. Alternatively, 
total correlations can be obtained from Eq.~(\ref{eq:pair-distribution-function}) via the test-particle 
route, where the two-particle density $\rho_{\nu\nu'}^{(2)}$ is determined by the additional calculation 
of the one-particle density profile around the first particle represented by an 
external field. Both routes would be consistent when the exact free energy functional was used. 
The compressibility route has advantages over the test-particle route when long-ranged mean-field 
Coulomb interactions are involved, whose direct correlations can be Fourier transformed analytically.

\subsection{Fundamental measure theory}

To calculate a total correlation function from the OZ relation in 
Eq.~(\ref{eq:ornstein-zernike-equation}), it is necessary to close it. A well-known example for such a 
closure is the Percus-Yevick approximation  
\begin{equation}
c^{(2)}(\vec{r}) \approx \left(1-\exp(\beta u(\vec{r}))\right)g^{(2)}(\vec{r})\,,
\end{equation}
where $u(\vec{r})$ is the pair interaction potential. For HSs, this approximation has been solved analytically by Wertheim \cite{wertheim_prl10_1963}. The results are included in Fig.~\ref{fig:Figure3}. 

In DFT, the direct correlations are explicitly given by Eq.~(\ref{eq:direct-correlations2}) via 
a second order functional derivative of the excess free energy functional. Thus, the 
OZ relation could be closed if the excess free energy 
functional ${\cal F}^{\rm exc}$ were known.
Unfortunately, the exact form of the functional is, in general, unknown \cite{mermin_prv137_1965}. However, many approximations 
exist. For hard particles and, especially, for HSs, FMT has been established as a quantitative benchmark theory \cite{oettel_pre86_2012}. 

In FMT \cite{rosenfeld_prl63_1989,roth_jpcm22_2010} the excess free energy
is expressed via the local excess free energy density $\Phi$, i.e.,
\begin{equation}
\beta {\cal F}^{\rm exc}[\{\rho_\nu\}] = \int_V \Phi(\vec{r}) d\vec{r} . 
\end{equation}
The function $\Phi$ is typically constructed to recover the correct Mayer $f$ function in the 
limit of low density such that the exact excess free energy is recovered in this 
limit \cite{rosenfeld_prl63_1989}. Extrapolation to higher densities leads to different versions of the FMT. 
Besides the original version of Rosenfeld \cite{rosenfeld_prl63_1989}, we mention, in particular, the 
extended deconvolution FMT for anisotropic convex-shaped hard particles 
\cite{hansen-goos_prl102_2009} and the White Bear and WBII versions 
\cite{roth_jpcm14_2002,yu_jcp117_2002,hansen-goos_jpcm18_2006} for HSs, which should include 
tensorial corrections to recover the exact zero-dimensional limit \cite{tarazona_prl84_2000}. 
Moreover, FMT can be derived from the virial series \cite{korden_pre85_2012,marechal_pre90_2014}. 
For our work we have chosen the WBII version 
with its tensorial correction, because it has been employed to accurately predict
not only the freezing transition in HS \cite{oettel_pre82_2010} but also phase coexistence 
and the involved crystal-fluid interface \cite{haertel_prl108_2012}. 
Its excess free energy density reads 
\begin{align}
\Phi(\vec{r}) =& -n_0\ln(1-n_3) \label{eq:local_excess_free_energy}\\
& + \left(1+\frac{1}{9}n_3^2\phi_2(n_3)\right) \frac{n_1n_2-\vec{n}_1\cdot\vec{n}_2}{1-n_3} \notag \\
& + \left(1-\frac{4}{9}n_3\phi_3(n_3)\right) \notag \\ & \times\frac{
n_2^3-3n_2\vec{n}_2\cdot\vec{n}_2 
+\tfrac{9}{2}\left(\vec{n}_2^T\cdot\tensor{n}_2\cdot\vec{n}_2
-{\rm tr}(\tensor{n}_{2}^3)\right)}{24\pi(1-n_3)^2} ,
\end{align}
where ${\rm tr}(\tensor{A})$ denotes the trace of the argument $\tensor{A}$ and the two functions $\phi_i(n_3)$ are
\begin{align}
\phi_2(n_3) =& \frac{6n_3-3n_3^2+6(1-n_3)\ln(1-n_3)}{n_3^3} , \\
\phi_3(n_3) =& \frac{6n_3-9n_3^2+6n_3^3+6(1-n_3)^2\ln(1-n_3)}{4n_3^3} \,,
\end{align}
with the so-called weighted densities $n_\alpha$. These weighted densities are given by the convolutions
\begin{equation}
n_\alpha(\vec{r})
=\sum_{\nu'=1}^{n} \int_{V} \rho_{\nu'}(\vec{r}{\,'}) w_{\nu'}^{(\alpha)}(\vec{r}-\vec{r}{\,'}) d\vec{r}{\,'}  \,.
\end{equation}
The convolutions weight the one-particle densities $\rho_{\nu'}$ of each species $\nu'$ with so-called 
weight functions $w_{\nu'}^{(\alpha)}$. The latter represent fundamental geometric measures 
like volume ($\alpha=3$ for three dimensions), surface area ($\alpha=2$ for two dimensions), mean diameter ($\alpha=1$ for one dimension), and curvature ($\alpha=0$ for zero dimensions) of a single-particle 
geometry.
For HS mixtures the weight functions of each species $\nu$ read 
\cite{rosenfeld_prl63_1989,tarazona_prl84_2000} 
\begin{align}
w_\nu^{(3)}(\vec{r}) &= \Theta(R_\nu-|\vec{r}|) \label{eq:weight3} , \\
w_\nu^{(2)}(\vec{r}) &= \delta(R_\nu-|\vec{r}|) , \\
w_\nu^{(1)}(\vec{r}) &= \frac{1}{4\pi R_\nu}\delta(R_\nu-|\vec{r}|) , \\
w_\nu^{(0)}(\vec{r}) &= \frac{1}{4\pi R_\nu^2}\delta(R_\nu-|\vec{r}|) , \\
\vec{w}_\nu^{(2)}(\vec{r}) &= \frac{\vec{r}}{|\vec{r}|}\delta(R_\nu-|\vec{r}|) , \\
\vec{w}_\nu^{(1)}(\vec{r}) &= \frac{\vec{r}}{|\vec{r}|}\frac{1}{4\pi R_\nu}\delta(R_\nu-|\vec{r}|) , \\
\tensor{w}_\nu^{(2)}(\vec{r}) &= 
 \left( \frac{\vec{r}\cdot\vec{r}^T}{|\vec{r}|^2} - \frac{\tensor{{\rm I}}}{3} \right) \delta(R_\nu-|\vec{r}|) , 
 \label{eq:weight2tensor}
\end{align}
where $R_{\nu}=\sigma_\nu/2$ denotes the radius of a sphere with diameter $\sigma_\nu$. 
Furthermore, the tensor product $\vec{r}\cdot\vec{r}^T$, the unit matrix $\tensor{{\rm I}}$, and the 
transposed $\vec{r}^T$ of a vector $\vec{r}$ have been used.

Via the framework of DFT the equation of state with pressure $p=-\Omega/V$ can be determined, as already exemplarily presented in 
Fig.~\ref{fig:Figure2} for a two-component HS mixture together 
with simulation results.

As for all FMT functionals with an excess free energy density that depends only on the weighted 
densities $n_{\alpha}$, the direct pair-correlation 
functions, as defined in Eq.~(\ref{eq:direct-correlations2}), are 
\begin{align}
& -c_{\nu\nu'}^{(2)}(\vec{r},\vec{r}{\,}') \label{eq:direct_correlation_function} \\
&  \quad = \sum_{\alpha,\beta} \int_{V} 
\frac{\partial^2\Phi}{\partial n_\alpha\partial n_\beta}(\vec{r}{\,}'') 
w_\nu^{(\alpha)}(\vec{r}{\,}''-\vec{r}) w_{\nu'}^{(\beta)}(\vec{r}{\,}''-\vec{r}{\,}') d\vec{r}{\,}'' . \notag
\end{align}
In bulk, the derivative with respect to the weighted densities becomes independent of 
the spatial coordinate and the direct correlation function can be calculated 
analytically \cite{rosenfeld_prl63_1989,roth_jpcm14_2002,oettel_pre86_2012}. 
For our anisotropic system, we report in the next section a semianalytical form 
for general multicomponent mixtures in the framework of FMT.

\subsection{Numerical details of FMT and OZ calculations in restricted geometries}

We approach the equilibrium density profiles by repeatedly applying Eq.~(\ref{eq:ele-density-profile}). 
During each iteration step $i$, the right-hand side of Eq.~(\ref{eq:ele-density-profile}) is applied to 
the actual set $\Gamma_i\equiv\{\rho_\nu\}_i$ of density profiles to achieve a new set $\Gamma_i^{\rm new}$ 
from the left-hand side of Eq.~(\ref{eq:ele-density-profile}). The new profiles are mixed with the actual 
ones to generate a set $\Gamma_{i+1}$ by adding a fraction $\alpha$ from the new ones in $\Gamma_i^{\rm new}$ 
and a fraction $1-\alpha$ from the recent ones in $\Gamma_i$. This procedure is repeated until the largest 
local deviation between all the new and the respective recent density profiles becomes smaller than a threshold $\epsilon$. 
We have started each Picard iteration from the bulk density profiles, where we simply neglect the wall. Typically after 
around 2500 iteration steps the profiles reached an accuracy of $\epsilon=10^{-6}$, while the mixing parameter 
changed from an initial value of $\alpha=10^{-8}$ to a final $\alpha=10^{-4}$ during the iteration. 

As a flat wall is introduced into the system, due to the symmetry of the structure close to that wall, all density profiles $\rho_\nu$ as well as all 
derivatives $\partial^2\Phi/(\partial n_\alpha\partial n_\beta)$ in 
Eq.~(\ref{eq:direct_correlation_function}) depend solely on 
the spatial coordinate $z$ perpendicular to the wall. Furthermore, 
the direct correlation functions depend only on three coordinates, i.e.,  
$c_{\nu\nu'}^{(2)}(r,z,z')$, as discussed in Sec.~\ref{sec:section2}. 

For numerical reasons, we sample our functions at a distance $L$ between the wall and the bulk fluid and at equidistant discrete points 
$z_i=i d_\text{z}$, with $d_\text{z}=L/M$ for $i=0, \ldots ,M-1$. When we consider intervals 
$I_i=[z_i-\tfrac{1}{2}d_\text{z},z_i+\tfrac{1}{2}d_\text{z}]$, we can split the integration 
volume $V=\mathcal{R}^3$ on the right-hand side of Eq.~(\ref{eq:direct_correlation_function}) 
into slices $V_i=\mathcal{R}^2\times I_i$ and rewrite the direct 
correlation functions as 
\begin{align}
-c_{\nu\nu'}^{(2)}(\vec{r},\vec{r}{\,}') 
\approx & \sum_{i=0}^{M-1} \sum_\alpha \sum_\beta 
\frac{\partial^2\Phi(z_i)}{\partial n_\alpha\partial n_\beta} \label{eq:c2:stepwise-constant} \\
& \times \int_{V_i} 
w_\nu^{(\alpha)}(\vec{r}{\,}''-\vec{r}) w_{\nu'}^{(\beta)}(\vec{r}{\,}''-\vec{r}{\,}') d\vec{r}{\,}'' .
\notag
\end{align}
In order to calculate the direct correlation functions, it is necessary 
to compute the integral in Eq.~(\ref{eq:c2:stepwise-constant}), which, for given combinations of particle species and weight functions, 
depends solely on the interval $I$ and the distance $\vec{\Delta}=\vec r{\,}' - \vec r$. Thus, we define auxiliary functions 
\begin{equation}
W_{\nu\nu'}^{(\alpha\beta)}(I, \vec{\Delta}) 
:= \int_{\mathcal{R}^2\times I} 
w_\nu^{(\alpha)}(\vec{r}{\,}'') 
w_{\nu'}^{(\beta)}(\vec{r}{\,}''-\vec{\Delta}) d\vec{r}{\,}'' ,
\label{eq:c2:define-WW}
\end{equation}
which we precompute analytically whenever possible. This reduces the 
computational cost significantly. For further details about the calculations 
of Eq.~(\ref{eq:c2:define-WW}) we refer to Appendix \ref{sec:appendix2}.

Finally, the knowledge of the density profiles $\rho_\nu$ and of the direct correlations 
$c_{\nu\nu'}^{(2)}$ enables us to determine the total correlations $h_{\nu\nu'}^{(2)}$ via the 
OZ relation from Eq.~(\ref{eq:ornstein-zernike-equation}). It is useful to solve this 
relation partially in Fourier space to exploit the symmetries of our system. 
For this purpose, we define an in-plane Fourier or 
Hankel transform (see Appendix \ref{sec:appendix1}) by 
\begin{align}
	& {\rm H}_{\rm r}\big\{h_{\nu\nu'}^{(2)}(\cdot,z,z')\big\}(K) 
\equiv h_{\nu\nu'}^{(2)}(K,z,z') \notag \\
%	= \frac{1}{2\pi}\int h_{\nu\nu'}^{(2)}(r,z,z') e^{-\imath \vec{K}\cdot\vec{r}} d\vec{r} , \notag
& \quad = \frac{1}{2\pi}\int_0^{\infty} \int_0^{2\pi} 
	r h_{\nu\nu'}^{(2)}(r,z,z') e^{-\imath K r \cos(\vartheta)} d\vartheta dr , 
	\label{eq:Hankel-transform}
\end{align}
which only assigns the radial components of a function and usually is employed to obtain structure 
factors of layers parallel to a symmetry-breaking wall (cf. \cite{tarazona_mph54_1985}).
With such a transform, the OZ relation 
from Eq.~(\ref{eq:ornstein-zernike-equation}) can be 
rewritten in the form %with respect to the cylindric geometry by 
\begin{align}
	& {\rm H}_{\rm r}\big\{ h_{\nu\nu'}^{(2)}(\cdot,z,z') \big\}(K) 
	\label{eq:ornstein-zernike-equation-cylindrical} \\
	=& {\rm H}_{\rm r}\big\{ c_{\nu\nu'}^{(2)}(\cdot,z,z') \big\}(K) \notag \\
	 & + 2\pi \sum_{\nu''=1}^{n} \int_{-\infty}^{\infty} \rho_{\nu''}(z'')  \notag \\
 & \quad \times \left[ {\rm H}_{\rm r}\big\{ h_{\nu\nu''}^{(2)}(\cdot,z,z'') \big\}%(|\vec{k}|)
{\rm H}_{\rm r}\big\{ c_{\nu''\nu'}^{(2)}(\cdot,z'',z') \big\} \right] (K) dz'' . \notag
\end{align}
For several values $K$, we determined the total correlations from this equation using an iterative numerical 
scheme (see also Appendix \ref{sec:appendix1}). 
In order to cope with numerical circumstances, we define our discrete lattice for the radial 
coordinate $r$ in a way that the value $r=0$ is avoided in real space. For this reason, in this work 
we solely provide data, where the radial component is very close but not equal to $0$.

%%%%%%%%%%%%%%%%%%%%%%%%%%%%%%%
%SECTION 5 %%%%%%%%%%%%%%%%%%%%
%%%%%%%%%%%%%%%%%%%%%%%%%%%%%%%
\section{Results}
\label{sec:section5}

In this section we quantitatively compare the results that we obtain from our multi-component DFT and the BD simulations.
First, we focus on one-particle densities. Second, we bear in mind the anisotropy in our system and consider the two-particle correlations. Consequently, all these 
results are employed in order to quantitatively analyze the contact properties of particles. These contact values are directly related to the anisotropic force distribution acting on a particle. As a result, the net force for a particle can be determined (cf. \cite{Nygard2014}). The nonuniform distribution of forces leads to the differences between effective 
diffusion coefficients in different directions. Finally, 
we demonstrate the impact of polydispersity by a comparison between our findings for a binary 
and a six-component mixture. In this context we discover a significant improvement in the agreement between the predictions of DFT calculations
and the results of BD simulations for an increasing number of particle species. 

\begin{figure*}
\centering 
\includegraphics[width=17.0cm]{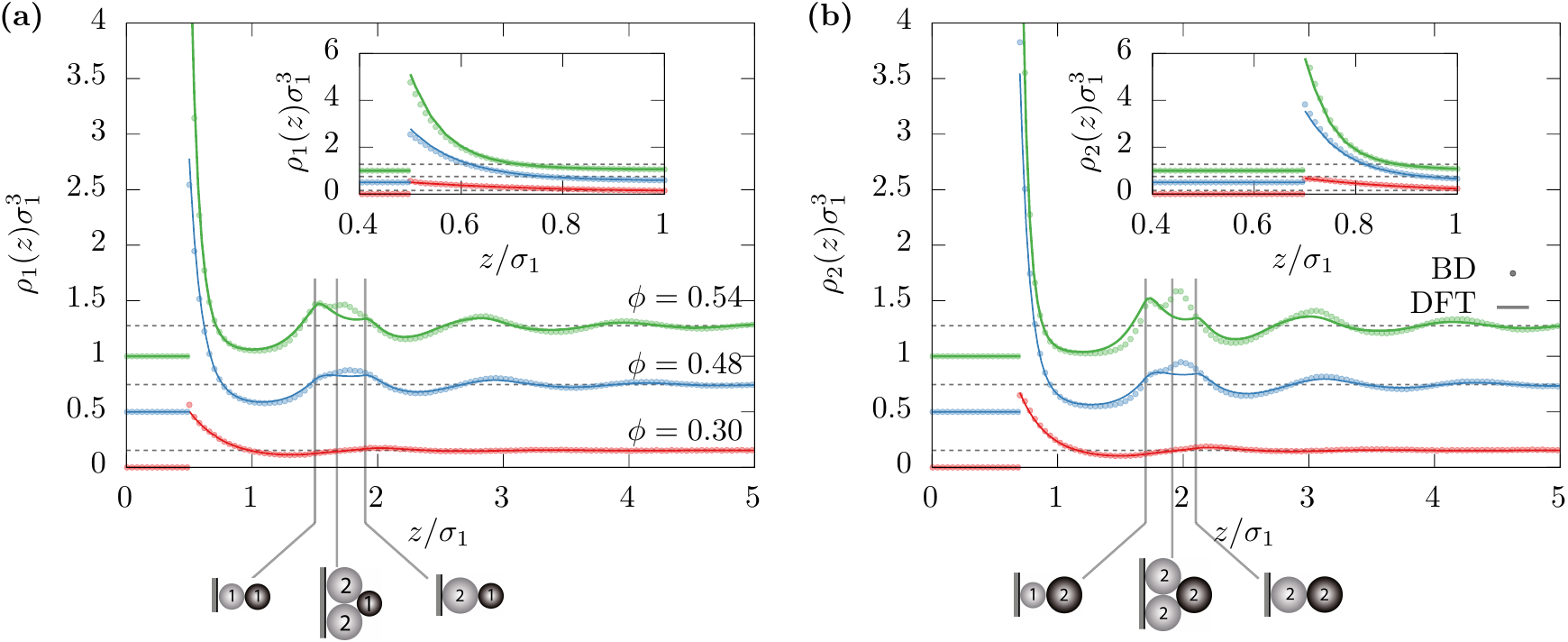}
\caption{\label{fig:Figure4}(Color online) 
Density profiles of (a) small and (b) large particles in binary (50:50) mixtures of HSs with diameters $\sigma_1$ and $\sigma_2=1.4\sigma_1$ in the vicinity of a flat hard wall (at $z=0$). 
Circles represent simulation data, whereas results of DFT calculations are represented by solid lines. 
To enhance readability, density profiles are shifted upward for different packing fractions by 
$0.5$ ($\phi=0.48$) and $1.0$ ($\phi=0.54$) and the dashed lines denote the bulk values. The small sketches at the bottom illustrate distinct packings 
of spheres. }
\end{figure*}

\subsection{One-particle density profiles}

In Fig.~\ref{fig:Figure4} we show density 
profiles of both DFT calculations and BD simulations for small and large particles in the binary (50:50) mixture of HS with diameters $\sigma_1$ and $\sigma_2=1.4\sigma_1$ as described in Sec. \ref{sec:section2}. The bulk densities have been fixed such that 
the corresponding total packing fractions are deep in the liquid phase ($\phi=0.3$), 
close to the fluid-crystal transition in monodisperse systems ($\phi=0.48$), and in the regime 
where glassy dynamics is observed ($\phi=0.54$). 
The most obvious differences between DFT calculations and BD simulation results occur in the second-layer 
peak of the density profiles. Especially in the profiles of higher bulk densities, the second-layer peak splits up into two peaks in the case of the simulation results (circles in Fig.~\ref{fig:Figure4}) or they just contain shoulders in the case of the DFT predictions (solid lines). Each local peak or shoulder can be connected by a particular stacking of particles belonging to different 
species, as illustrated by the sketches at the bottom of Fig.~\ref{fig:Figure4}. 
Note that local crystal-like ordering is not precisely captured in our DFT approach because we assume translational invariance along the wall. As a consequence, as soon as such locally ordered structures are preferred by the system, our DFT predictions become less accurate, 
even though the overall structure is not yet a crystal. 
Accordingly, the overall agreement between simulations and theory is very good
for low packing fractions.

\begin{figure}
\centering
\includegraphics[width=8.3cm]{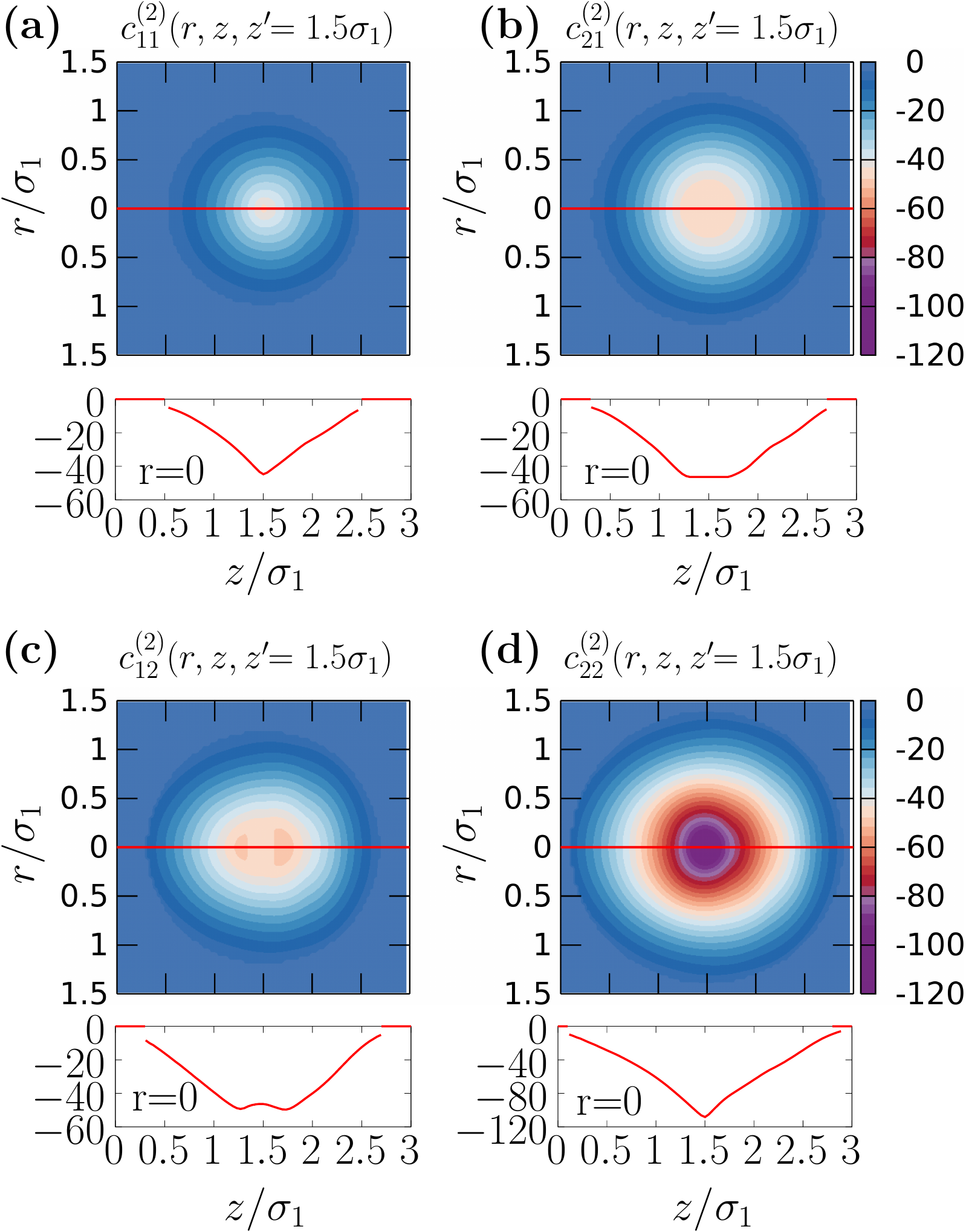}
\caption{\label{fig:Figure5}(Color online) 
Direct correlation functions obtained from DFT using Eqs.~(\ref{eq:c2:stepwise-constant}) 
and (\ref{eq:c2:define-WW}) for the binary HS mixture as explained in the text. The reference particle is fixed at $z'=1.5\sigma_1$. 
For a second particle at position $r,z$ we show the direct 
correlations (a) $c_{11}^{(2)}$, between small and small, (b) $c_{21}^{(2)}$, between large and small, (c) $c_{12}^{(2)}$, between small and large, 
and (d) $c_{22}^{(2)}$, between large and large particles. Note that the second index always denotes the fixed reference particle. The total volume fraction is $\phi=0.5$. 
Below the contour plots the profiles along the $z$ axis with $r=0$ are shown [represented by solid (red) lines in the contour plots]. }
\end{figure}

\begin{figure}
\centering
\includegraphics[width=8.3cm]{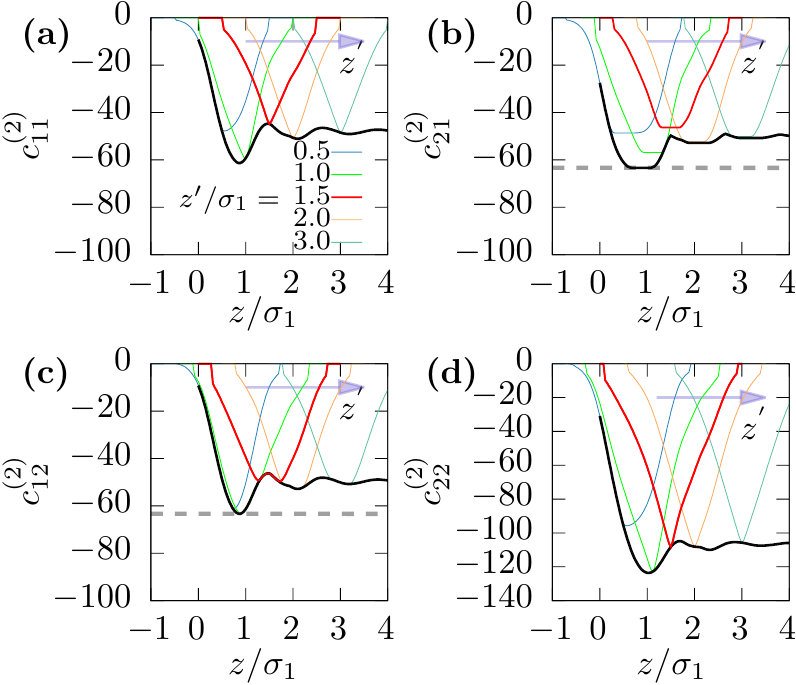}
\caption{\label{fig:Figure6}(Color online) 
Direct correlation profiles along the $z$ axis as shown in Fig.~\ref{fig:Figure5}, but for various positions $z'$ of the reference particle. 
The profiles from Fig.~\ref{fig:Figure5} with $z'=1.5\sigma_1$ are shown by solid bold lines. Again, the correlations are between (a) small and small, (b) large and small, (c) small and large, 
and (d) large and large particles. In addition, 
the envelopes of all profiles are shown. 
Dashed horizontal lines show the positions of the minima in (b) and (c), which are equal.}
\end{figure}

\begin{figure*}
\centering
\includegraphics[width=17.0cm]{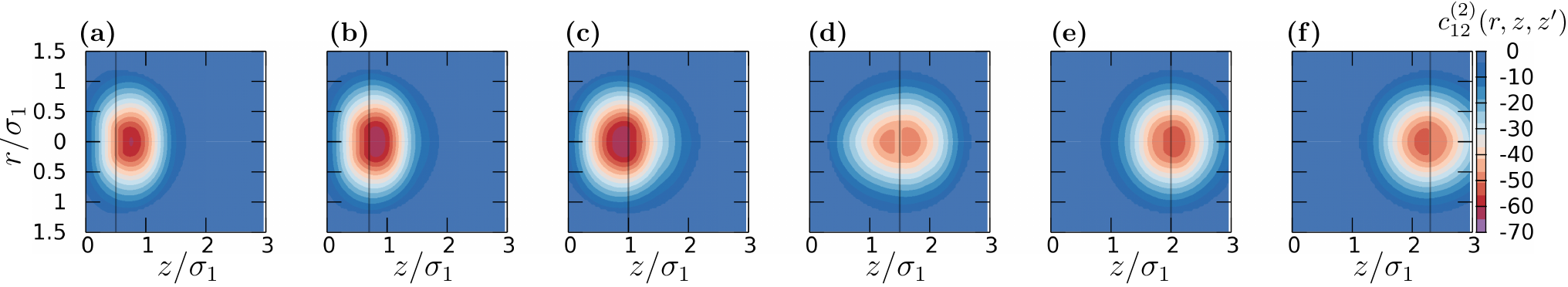}
\caption{\label{fig:Figure7}(Color online) 
Direct correlation functions $c_{12}^{(2)}(r,z,z')$ as shown in 
Fig.~\ref{fig:Figure5}(c) for a large reference particle at different positions $z'$, which are marked by vertical lines at (a) $0.5$, (b) $0.7$, (c) $1.0$, (d) $1.5$, 
(e) $2.0$, and (f) $2.3$. }
\end{figure*}

\subsection{Two-particle correlations}

In DFT, the two-particle or pair correlations can be obtained via the test-particle or the 
compressibility route. For the first, density profiles are determined around a fixed test 
particle which results in an effective two-particle density. We follow 
the compressibility route, where the direct correlation functions $c_{\nu\nu'}^{(2)}$ from DFT 
are used to close the OZ relation from Eq.~(\ref{eq:ornstein-zernike-equation}). 
Using the WBII functional, we obtain the density profiles $\rho_\nu$ and 
direct correlation functions $c_{\nu\nu'}^{(2)}$, where we calculate the latter directly 
via Eqs.~(\ref{eq:c2:stepwise-constant}) and (\ref{eq:c2:define-WW}) for our 
inhomogeneous system. 
An advantage of the compressibility route over the test-particle route is that no boundary 
effects in direction $r$ parallel to the wall are involved in the calculation of direct correlations. 
Moreover, the latter are short-ranged for HSs and can be Fourier transformed numerically for 
arbitrary sets of vectors $\vec{k}$ in Fourier space. Thus, the full structure factor $S(\vec{k})$ is 
attainable without restrictions resulting from a finite extension in the $r$ direction.

\subsubsection{Direct correlations}

The direct correlations are shown exemplarily in 
Figs.~\ref{fig:Figure5}-\ref{fig:Figure7} for the binary
mixture of HSs. 
First, in Fig.~\ref{fig:Figure5}, we compare the $c_{\nu\nu'}^{(2)}$ values for the 
four combinations between the two species (small-small, large-small, small-large, 
and large-large). The position of the reference particle is fixed at $z'=1.5\sigma_1$ and the direct correlations are plotted as functions of the position of the other particle, where the position is expressed in the natural cylindrical coordinates $(r,z)$. In addition, we 
show the profile along the $z$ axis together with each plot. While the correlations between 
two large or two small particles differ only by a constant factor and by the length scale, 
the correlations between a small and a large particle depend on which particle is used as the 
reference particle. In both cases the direct correlation functions do not have one clear 
minimum. While in the case of a small reference particle there is a plateau with an extent of $0.4\sigma_1$ in the $z$ direction, in the case of 
a large reference particle there are two distinct minima, at $z\approx 1.3\sigma_1$ and $z\approx 1.7\sigma_1$. Note that in bulk, both correlation 
functions between large and small particles are identical [see Fig.~\ref{fig:Figure3}(a)] and possess a plateau for $r<0.2\sigma_1$ where the direct correlation function is 
constant. The plateau is due to the fact that the intersection volume of the two spheres does not change as long as the small 
particle is located completely inside the large one as sketched in the inset in Fig.~\ref{fig:Figure3}(a). Therefore, the value of the integral in 
Eq.~(\ref{eq:direct_correlation_function}) does not change and the observed plateau develops. 
Back to the anisotropic case in Fig.~\ref{fig:Figure5}, a similar explanation 
holds: When the position of a small particle is fixed, as in Fig.~\ref{fig:Figure5}(b), 
the integration volume $V$ in Eq.~(\ref{eq:direct_correlation_function}) is restricted 
to the shape of this particle as long as the small particle is completely contained 
inside the larger one. In contrast, when a large particle is fixed, as in 
Fig.~\ref{fig:Figure5}(c), the previously mentioned integration volume 
depends on the position of the small particle. Therefore, the result of the integral in Eq.~(\ref{eq:direct_correlation_function}) 
depends on the relative positions of the particles via the derivative of the excess free energy density $\Phi$. The resulting 
direct correlation function is similar to the self-correlations between two small particles, because 
the relevant combinations of weight functions $w^{(\alpha)}$ that enter 
Eq.~(\ref{eq:direct_correlation_function}) give the same results in this case (for further details 
see Appendix \ref{sec:app-case1}, case 3). \label{sec:text-case1}

In Fig.~\ref{fig:Figure6} we compare slice cuts of the direct correlation profiles along the $z$ axis 
for various positions $z'$ of the reference particle. Additionally, we draw the 
envelope to all shown profiles. Figures~\ref{fig:Figure6}(a) and \ref{fig:Figure6}(c) demonstrate the similarity between 
small-small and small-large correlations, mentioned in the previous paragraph.

In Figs.~\ref{fig:Figure5}(c) and \ref{fig:Figure6}(c), we observe a splitting of the minimum of the direct 
correlation function into two minima. The splitting occurs for the parameters where the direct correlation functions reach a local maximum in the corresponding envelope 
of the profiles as can be seen in Fig.~\ref{fig:Figure6}(c). This suggests that there exists 
a $z$-dependent maximal correlation for a particular combination of species. In Fig.~\ref{fig:Figure7} we show a series of direct correlation functions with varying position $z'$ of the reference particle. 
These positions are marked by vertical lines, and obviously, the absolute minimum of the direct correlations is located in the vicinity of these positions. Specifically, 
the global minimum of the direct correlation functions shown in Fig.~\ref{fig:Figure7} can be found at $z>z'$ in Figs.~\ref{fig:Figure7}(a), \ref{fig:Figure7}(b), and \ref{fig:Figure7}(e) but 
at $z<z'$ in Figs.~\ref{fig:Figure7}(c) and \ref{fig:Figure7}(f). In Fig.~\ref{fig:Figure7}(d) the minimum is split into two local minima on both sides of the center of the reference particle. 
This behavior can again be understood from studying the corresponding profiles in 
Fig.~\ref{fig:Figure6}(c), where the shape of the region around the minimum of each profile always follows the maximal possible correlation, given by the envelope. The anisotropic arrangement of the direct correlation functions around the center of the reference particle will lead to anisotropic forces as we show later.

\subsubsection{Total correlations}

\begin{figure}
\centering
\includegraphics[width=8.0cm]{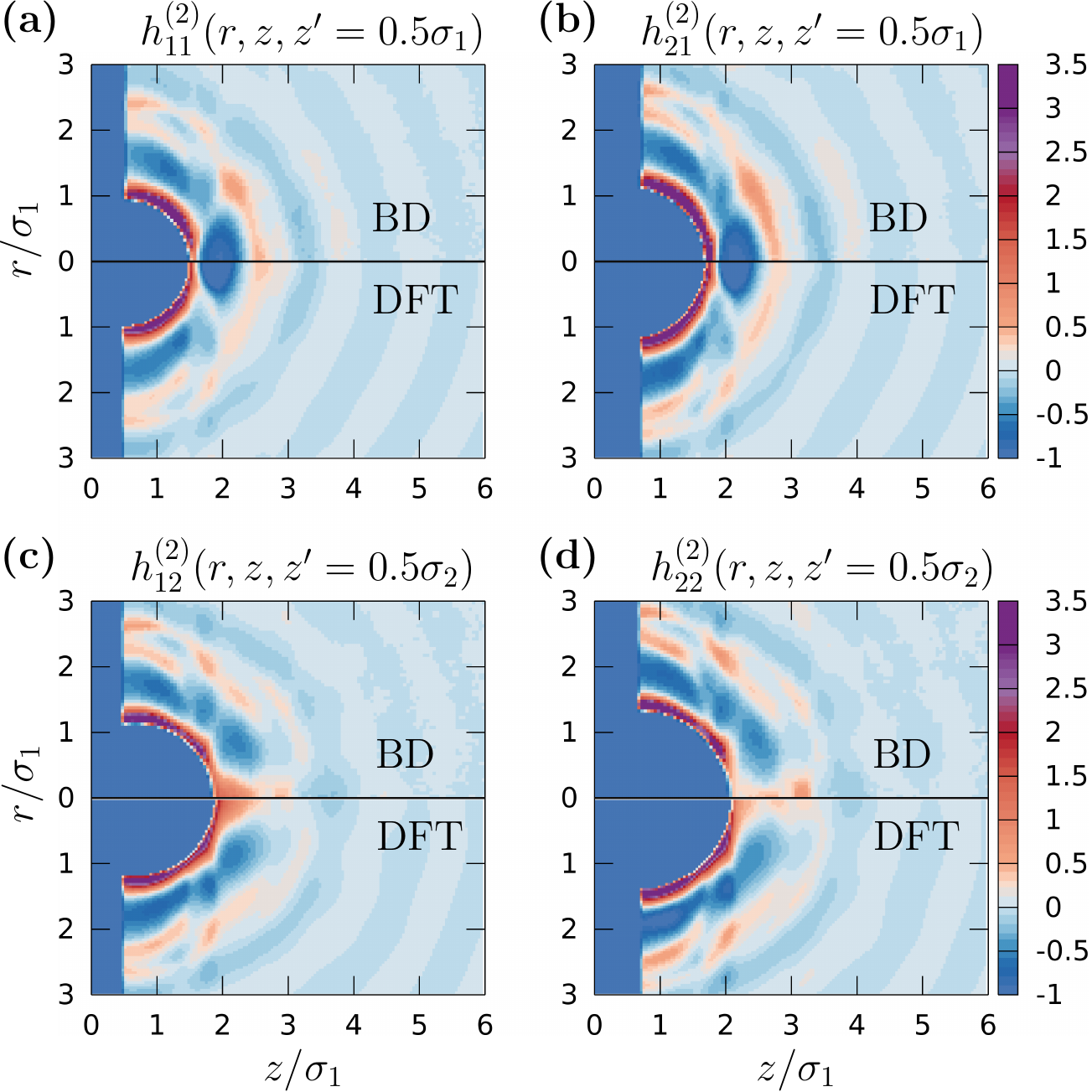}
\caption{\label{fig:Figure8}(Color online) 
Total correlation functions $h_{\nu\nu'}^{(2)}(r,z,z')$ for a reference particle at position 
$z'=0.5\sigma_{\nu'}$ between (a) small and small, (b) large and small, (c) small and large, 
and (d) large and large particles, where the second particle denotes the reference particle. The packing fraction is $\phi=0.5$ in the bulk limit and each plot is split 
up into data from Brownian dynamics (BD) simulations (top) and DFT results (bottom), where the total correlation functions were determined via the OZ relation. 
Note that in the case of the DFT calculations all numerical artifacts at forbidden positions (inside the wall and inside the reference particle) have been reset to $-1$.} 
\end{figure}  

\begin{figure*}
\centering
\includegraphics[width=17.0cm]{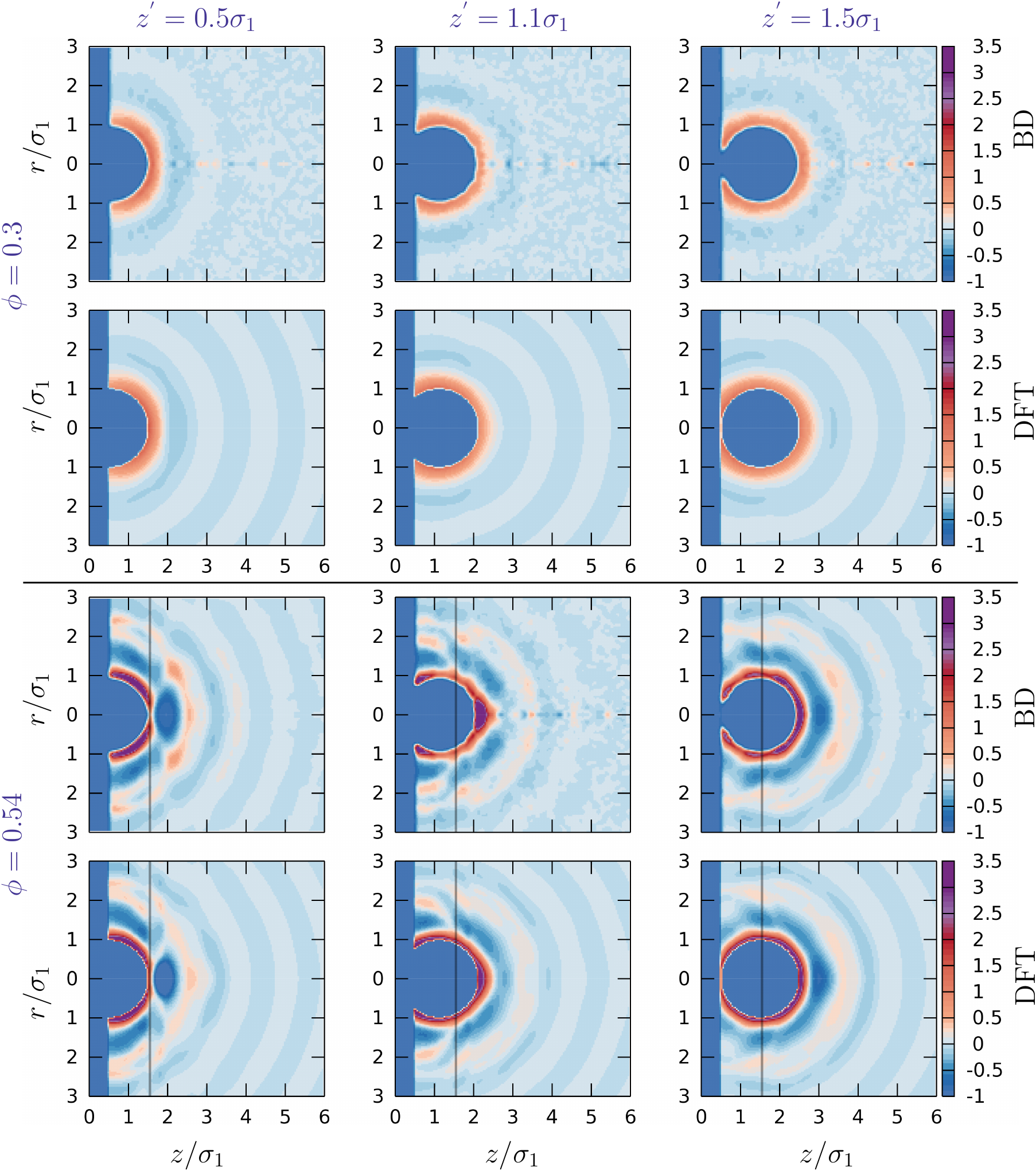}
\caption{\label{fig:Figure9}(Color online) 
Total correlations $h_{11}^{(2)}(r,z,z')$ between small and small particles with diameter
$\sigma_1$. The reference 
particle is fixed at a position $z'$ and the position of the other particle is given in cylindrical coordinates $(r,z)$. The hard wall is located at $z=0$. 
Brownian dynamics (BD) data are shown in the first and third rows; DFT results, in the second and forth row. In the top two rows a low packing fraction, $\phi=0.3$, is employed, while a high-density case, with $\phi=0.54$, 
is shown in the bottom two rows. Each column denotes a different position $z'$ of the reference particle ($z'/\sigma_1=0.5,\,1.1,$ or $1.5$). All numerical artifacts at forbidden positions have been removed and reset to $-1$ in case of DFT results. The speckled pattern in the lower density simulation data arises from poorer statistics at the location of local minima in the density close to the $z$ axis with $r=0$. Vertical lines in the bottom rows 
indicate the positions of the profiles shown in Fig.~\ref{fig:Figure10}. }
\end{figure*} 

\begin{figure}
\centering
\includegraphics[width=8.0cm]{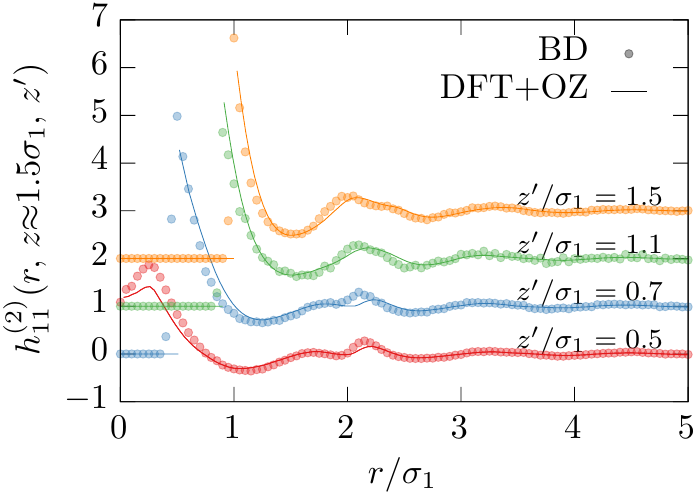}
\caption{\label{fig:Figure10}(Color online) 
Profiles of the total correlation function $h_{11}^{(2)}$ along the vertical lines drawn in Fig.~\ref{fig:Figure9} and located at the second layer of particles ($z=1.55\sigma_1$). 
The packing fraction of the system is $\phi=0.54$ and the slice cuts are shown for three positions $z'$ 
of the fixed reference particle. 
Solid lines denote DFT data and circles represent results from the respective Brownian dynamics (BD) simulations. 
For DFT all total correlations at forbidden positions have been reset to $-1$ and the curves have been shifted by 0.0, 1.0, 2.0, and 3.0 (from bottom to top).}
\end{figure} 

Starting from the direct correlations and one-particle densities determined with DFT, 
we calculate the total correlations between two particles using the OZ relation 
from Eq.~(\ref{eq:ornstein-zernike-equation-cylindrical}). As mentioned in 
Sec.~\ref{sec:OZ-relation}, this equation is exact, but we have to deal with 
numerics in order to perform this transformation. Especially, the finite number 
of Fourier modes in our discretization gives rise to artifacts. As we can see 
in Fig.~\ref{fig:Figure3}(b) for a bulk fluid, the resulting 
total correlation functions show unphysical values differing from $-1$ inside the 
core. Note that this behavior not only originates from numerical 
inconveniences during solving the OZ relation but also depends on the inconsistency of 
the approximate excess free energy functional we have used. Such inconsistencies 
are common for all approximate functionals and 
can only be resolved by the exact functional, which in general is not known 
\cite{mermin_prv137_1965}. In our case the specific artifacts in the forbidden regions 
could be avoided by employing the earlier-mentioned test-particle route via the 
two-particle density in Eq.~(\ref{eq:pair-distribution-function}), which does not show the 
deviations from $-1$ in forbidden regions, per definition. However, this route is expected to show deviations in other regions of the profiles where the 
compressibility route might work more precisely, because the forced hard potential of 
the test particle is not consistent with the properties of the approximate functional, 
e.g., increased correlations in the particle core. 

Similarly to Fig.~\ref{fig:Figure5}, we show the total correlation functions for all 
possible pairs of particles in Fig.~\ref{fig:Figure8}. In addition to our results 
determined with DFT and the OZ relation, we plot the total correlations obtained from BD 
simulations via the test-particle route, which is natural for simulations. 
Simulation results are presented at the top of each 
plot; at the bottom the immediate comparison to the DFT results is shown. 
In general, both DFT calculations and BD simulations show a good agreement for all total particle 
correlations. However, as noted in the case of the direct correlation function 
in the previous subsection, the corresponding local structures are usually underestimated by 
DFT predictions whenever local ordering occurs. For example, deviations can be seen in 
Figs.~\ref{fig:Figure8}(a) and \ref{fig:Figure8}(b), where simulations lead to stronger correlations between 
the fixed reference particle and a second particle at $(r\approx 1\sigma_1,\,z\approx 2.1\sigma_1)$. In this position, particles in the second 
layer of a local fcc or bcc structure are located. Such orderings  occur more often for higher 
packing fractions and they are not incorporated in our DFT approach. 

In Fig.~\ref{fig:Figure9} a small reference particle is fixed at different positions $z'$ and the total correlations with another small particle at position $(r,z)$ are shown. Besides the previously discussed small deviations, the comparison between DFT calculations and BD simulations in general reveals 
a good quantitative agreement.

In order to study possible deviations in more detail, we show the profiles along 
the vertical lines in the bottom rows in Fig.~\ref{fig:Figure9} separately 
in Fig.~\ref{fig:Figure10}. Note that these data are taken at the rather high packing fraction 
$\phi=0.54$, where glassy dynamics sets in. 
Nevertheless, the overall agreement is still good. The most pronounced differences occur 
close to particle contact. In the simulation data this behavior is affected by two 
effects: on one side, the slight softness of the repulsive interactions and, on the other 
side, the uncertainty of the actual position of the reference particle due to the 
discretization of the $z$ axis. In the next subsection, we study contact values and 
resulting forces on the test particle in more detail.

\subsection{Contact values and anisotropic forces}

\begin{figure*}
\centering
\includegraphics[width=16.0cm]{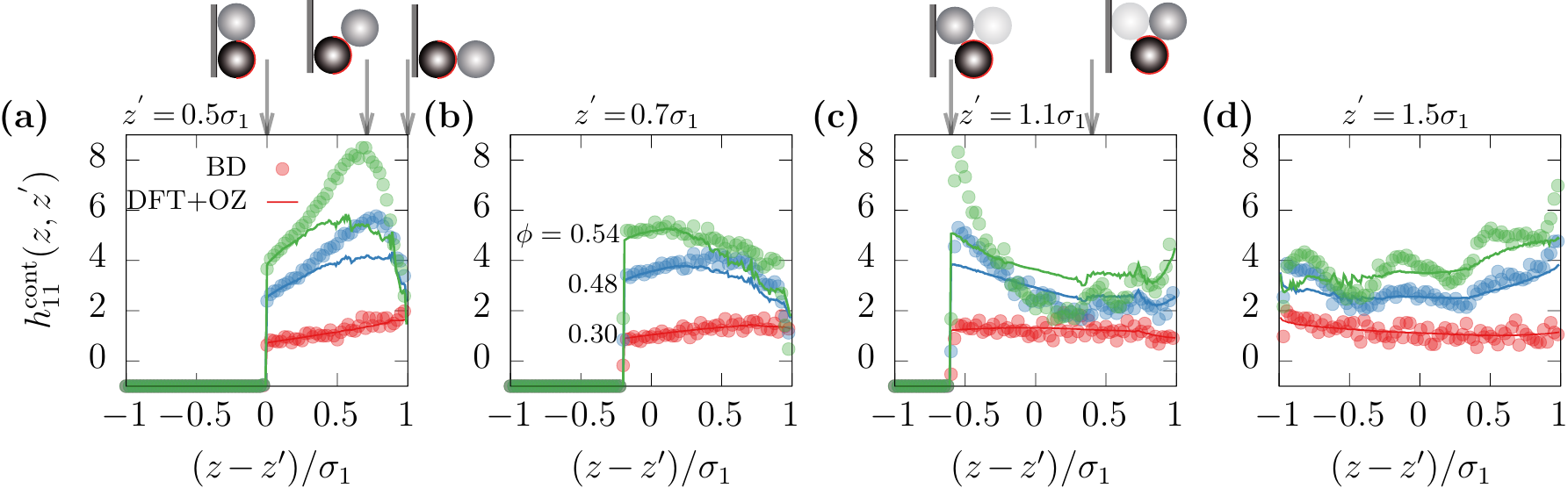}
\caption{\label{fig:Figure11}(Color online) 
The total correlation function $h^\text{cont}_{11}(z,z')$ at contact of a small reference particle 
at position $z'$ with a small neighbor particle at wall distance $z$. The position of the reference particle $z'$ is (a) $0.5\sigma_1$, (b) $0.7\sigma_1$, (c) $1.1\sigma_1$, and (d) $1.5\sigma_1$. 
Data from Brownian dynamics (BD) simulations (circles) and DFT results (lines) are shown for packing fractions 
$\phi=0.30$ (bottom curves), $0.48$ (middle curves), and $0.54$ (upper curves), as marked in (b). Sketches at the top illustrate certain arrangements of neighbor particles with the respective $(z-z')$ positions.
}
\end{figure*}

Anisotropy in structure results in an anisotropic distribution of forces acting on a particle. 
Obviously, such an anisotropic distribution can result in a nonvanishing net force. The force 
distribution and the net force depend on the total pair correlations at particle-particle contact. 
For this reason, we explored the value of the total pair-correlation functions 
$h_{\nu\nu'}^\text{cont}=g_{\nu\nu'}^\text{cont}-1$ 
at particle-particle contact. Note that the condition of 
contact effectively reduces the amount of independent parameters by one, i.e., $(z,z')$ instead of $(r,z,z')$. 

In Fig.~\ref{fig:Figure11} we present the contact values along the surface of a small reference 
particle in a binary mixture, which is located at several distances from the wall. Starting in 
Fig.~\ref{fig:Figure11}(a) with wall contact, the reference particle is slowly detached from the 
first layer at the wall in Figs.~\ref{fig:Figure11}(b) and (c) until it reaches the second layer 
in Fig.~\ref{fig:Figure11}(d). For these different positions, we compare results obtained from 
our BD simulations with the results calculated from DFT and the OZ relation. We find reasonable 
overall agreement. However, aside from statistical noise, some details of the data reveal significant 
differences: First, in Fig.~\ref{fig:Figure11}(a) the total correlations $h_{\nu\nu'}^\text{cont}(z,z')$ 
obtained from the simulations exhibit a very pronounced maximum at around $(z-z')/\sigma_1\approx0.71$ 
in the case of the two systems with higher densities. The contact values obtained from DFT also possess 
maxima at these positions, but they are less pronounced. Probably, this is again due to the neglect 
of local structure parallel to the wall in our theory. Indeed, the simulation data show some 
entropically favored contact correlations 
which are most obvious by the stronger oscillations in Fig.~\ref{fig:Figure11}(d). 

As mentioned before, anisotropies in structure also cause anisotropic force distributions. 
To determine these forces, we first consider a reduced Helmholtz free energy, which depends only 
on one so-called {\it reaction coordinate}. Typically, such a reduced free energy 
is achieved from the free energy of a multiparticle ensemble by integrating out all 
coordinates except for the reaction coordinate \cite{Trzesniak2007}. Then this 
coordinate can be used to describe transitions and reactions within a statistical 
manner \cite{Hanggi1990}. In our case, we want to disassemble the force on a single particle 
in the presence of a flat wall, 
where layers of particles form. In this situation, the natural choice for the reaction coordinate 
is the $z$ coordinate of a considered test particle of species $\nu$, such that the reduced free 
energy can be written as \cite{Trzesniak2007} 
\begin{equation}
F_\nu^{\rm red}(z) = -k_{\rm B}T \ln\big(\rho_\nu(z)\big) - k_{\rm B}T \ln(\Sigma) . 
\label{eq:potential_of_mean_force}
\end{equation}
The second term on the right-hand side of Eq.~(\ref{eq:potential_of_mean_force}) incorporates 
the partition function $\Sigma$ of the thermodynamic system but does not depend on $z$. 
Typically, Eq.~(\ref{eq:potential_of_mean_force}) is called the potential of mean force and it can 
be connected formally to the mean force $f_{\nu,{\rm z}}(z)$ in the $z$ direction by a derivative with 
respect to the reaction coordinate $z$; i.e.,
\begin{equation}
f_{\nu,{\rm z}}(z) = - \frac{\partial F_\nu^{\rm red}(z) }{\partial z} 
= k_\text{B}T\frac{\partial \ln\big(\rho_\nu(z)\big)}{\partial z} . 
\label{eq:mean_force}
\end{equation}
Now the Lovett-Mou-Buff-Wertheim
equations \cite{Lovett1976,Wertheim1976} can be 
used to connect the gradient of the density profile, and therefore the resulting mean force, 
with the two-particle direct correlations by 
\begin{equation}
f_{\nu',{\rm z}}(z') = 2\pi k_{\rm B}T \sum_{\nu=1}^{n}\int 
r\, c^{(2)}_{\nu\nu'}(r,z,z')\frac{\partial \rho_{\nu}(z)}{\partial z} dz\, dr .
\label{eq:LMBW-from-c} 
\end{equation}
Via an orthogonality relation for the density-density correlations, 
which is a statement of the OZ relation \cite{tarazona_mph54_1985,hansen_book_2013}, 
this mean force can also be connected with the pair-correlation functions, leading to \cite{Lovett1976} 
\begin{equation}
	f_{\nu',{\rm z}}(z') = 2\pi k_{\rm B}T \sum_{\nu=1}^{n}\sigma_{\nu\nu'}\hspace*{-0.1cm}\int 
\rho_{\nu}(z)g^\text{cont}_{\nu\nu'}(z,z')\frac{z'-z}{\sigma_{\nu\nu'}}dz . 
\label{eq:LMBW-from-g}
\end{equation}
Note that Eqs.~(\ref{eq:mean_force}) and (\ref{eq:LMBW-from-g}) provide an exact relation between 
a one- and a two-particle correlation, 
because Eq.~(\ref{eq:LMBW-from-g}) corresponds to the first member of the Born-Green-Yvon 
hierarchy \cite{Kjellander1991,Nygard2014}. 

\begin{figure}
\centering
\includegraphics[width=8.0cm]{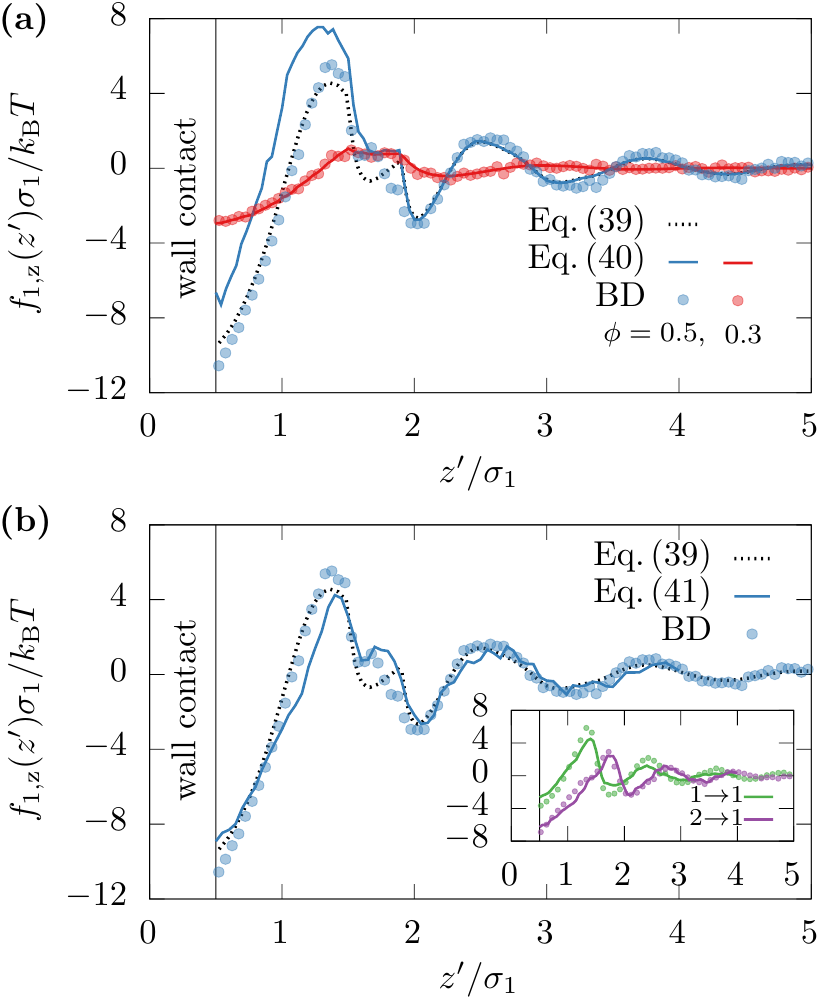}
\caption{\label{fig:Figure12}(Color online) 
Normalized force $f_{1,{\rm z}}$ on a small test particle at position $z'$ in the $z$ direction, which 
originates from the surrounding particles in a binary mixture at packing fractions $\phi=0.5$ (blue line) or $0.3$ (red line). Data from Brownian dynamics (BD) simulations 
are plotted with circles. For comparison, different methods to predict the force distribution from DFT are shown: Employing the potential of mean force as in Eq.~(\ref{eq:mean_force}) 
leads to the dotted lines, the solid lines in (a) denote the force distribution calculated from the direct correlations as in Eq.~(\ref{eq:LMBW-from-c}) for both packing 
fractions ($\phi=0.5$ and $0.3$), and the solid line in (b) corresponds to the distribution determined via the pair correlations as in Eq.~(\ref{eq:LMBW-from-g}) for the larger packing fraction ($\phi=0.5$) only. 
Inset in (b): Separated contributions from small ($1\to 1$) and large ($2\to 1$) particles to the mean force at $\phi=0.5$. }
\end{figure}

In Fig.~\ref{fig:Figure12}(a), we plot the net forces obtained from 
our theoretical calculations via Eqs.~(\ref{eq:mean_force}) and (\ref{eq:LMBW-from-c}); in Fig.~\ref{fig:Figure12}(b), we compare the results of Eqs.~(\ref{eq:mean_force})
and (\ref{eq:LMBW-from-g}).  In both figures, we additionally plot the forces directly obtained from 
our BD simulations for comparison. Clearly, the net forces that are theoretically obtained via the density profiles as in Eq.~(\ref{eq:mean_force}) match the simulation results very well. However, at high densities 
we observe a significant deviation between the curves at around $z=1.9\sigma_1$, where the small test particle can stack exactly on top of one large particle that is in contact with the wall and where local ordering 
might have a pronounced influence on the particles structure.
Employing Eqs.~(\ref{eq:LMBW-from-c}) and (\ref{eq:LMBW-from-g}) leads to forces that deviate from the simulation results for $z<1.9\sigma_1$. These differences are probably due to the thermodynamic 
inconsistency of the functional, which, for example, manifests in the differences between the 
compressibility and the test-particle route. Note that Eq.~(\ref{eq:mean_force}) corresponds to the test-particle route, because it solely involves the density profiles, 
while Eqs.~(\ref{eq:LMBW-from-c}) and (\ref{eq:LMBW-from-g}) involve the direct correlations. 
The latter seem to capture the behavior around $z=1.9\sigma_1$ 
better, while the results from Eq.~(\ref{eq:mean_force}) have a better agreement 
close to the wall.

Besides numerical inaccuracies, Eqs.~(\ref{eq:LMBW-from-c}) and (\ref{eq:LMBW-from-g}), in principle, are equivalent. Note, however, that only Eq.~(\ref{eq:LMBW-from-g}), 
where the forces are calculated using the pair correlations, offers direct access to the specific contributions of each particle species to the directional distribution of the net force. 
Such species-resolved contributions are shown in the inset 
in Fig.~(\ref{fig:Figure12})(b). In order to obtain this information from Eq.~(\ref{eq:LMBW-from-c}), where the forces depend on the direct correlation functions, one first has to determine the 
impact of one particle on another by integrating over all possible amounts of intermediate particles.

The results in the inset in Fig.~\ref{fig:Figure12}(b) show that, close to the wall, the 
large particles push the small test particle more strongly to the wall than the small particles do. 
If the test particle is moved away from the wall, first the contribution from the small particles 
reverses its direction such that they start pushing the particle away from the wall. For the larger particles the reversal of force direction occurs at a larger distance from the wall. Between the positions of these two reversals of directions, the resulting net force is small.

\subsection{Comparison among one-, two-, and six-component mixtures}

\begin{figure}
\centering
\includegraphics[width=8.3cm]{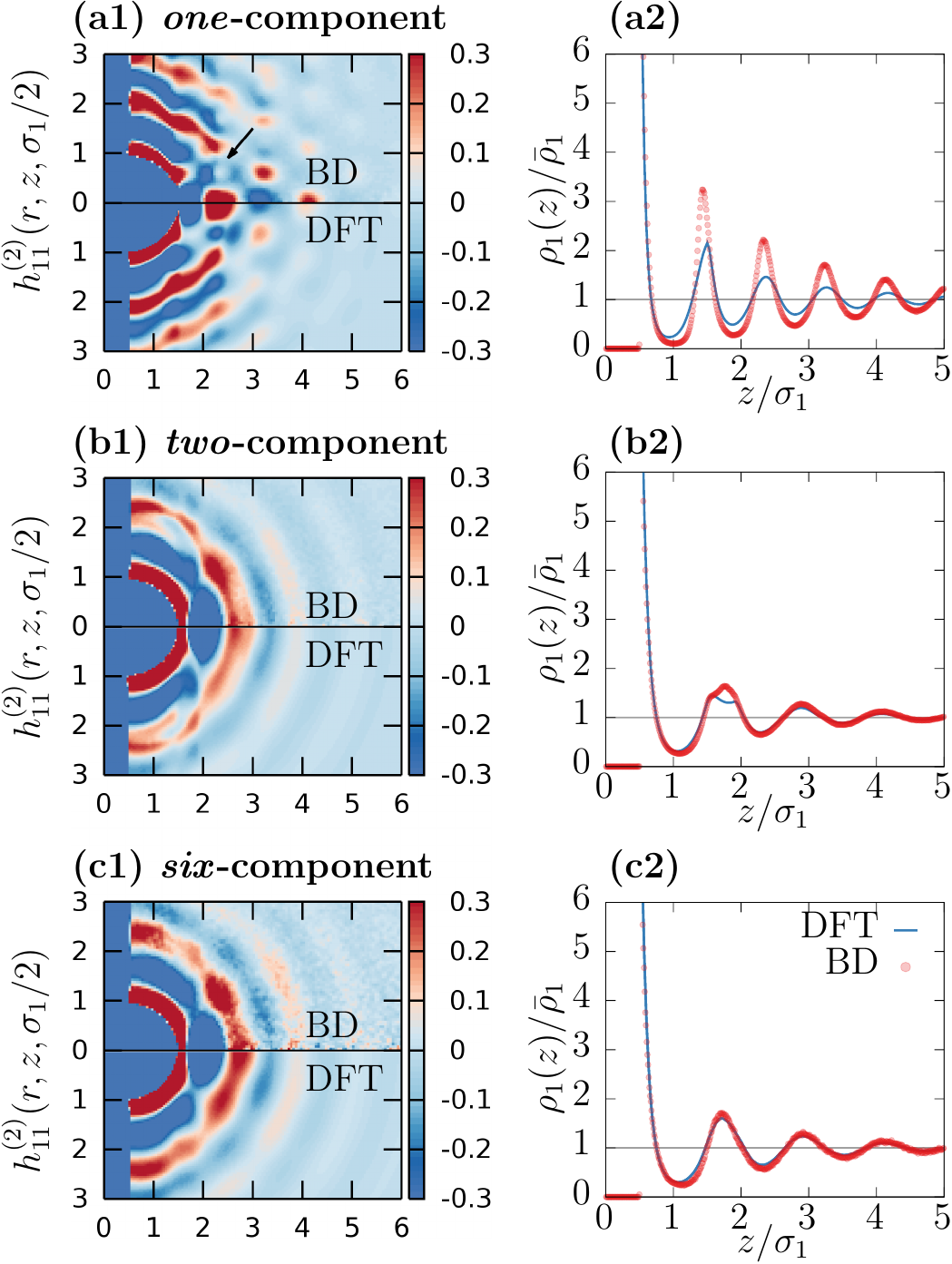}
\caption{\label{fig:Figure13}(Color online) 
Two- and one-particle correlations at a packing fraction of $\phi=0.5$ for 
(a) a \emph{one}-component, (b) a \emph{two}-component, and (c) a \emph{six}-component mixture, 
where the distribution of particle diameters is explained in the text. All plots compare data 
from our BD simulations and DFT. (a1, b1, c1) Total self-correlation 
functions $h_{11}^{(2)}(r,z,\sigma_1/2)$; 
(a2, b2, c2) accompanying density profiles $\rho_1(z)$. The latter are shown 
normalized with the respective bulk density $\bar{\rho}_1$. }
\end{figure} 

For an increasing number of components in a mixture, local ordering is suppressed even at high 
densities. As we show in the following the signatures of local structures in one- or 
two-particle correlations are smeared out with an increasing number of components. As a 
consequence, DFT calculations that neglect some types of local ordering become more accurate 
for such an increasing number of components. 

In Fig.~\ref{fig:Figure13} we demonstrate this effect for a packing fraction of 
$\phi=0.5$, where a one-component [Fig.~\ref{fig:Figure13}(a)], a two-component [Fig.~\ref{fig:Figure13}(b)], and a six-component [Fig.~\ref{fig:Figure13}(c)] system 
have been used. The binary mixture is the same as discussed throughout this work, 
with particle diameters $\sigma_1$ and $\sigma_2=1.4\sigma_1$, while the multicomponent 
system contains an equimolar mixture with particles of six discrete sizes: 
$\sigma_1$, $1.1\sigma_1$, $1.2\sigma_1$, $1.3\sigma_1$, $1.4\sigma_1$, and $1.5\sigma_1$. 
In Figs.~\ref{fig:Figure13}(a1), \ref{fig:Figure13}(b1), and \ref{fig:Figure13}(c1) we show the total self-correlation 
function $h_{11}^{(2)}(r,z,z')$ of the smallest particles, where one particle is in contact with the wall. 
Obviously, for the monodisperse case the peaks are very pronounced, and due to the high 
packing fraction of $\phi=0.5$ and the induced anisotropy, crystal-like structures are 
visible already on the two-particle level. As expected, major differences occur between 
DFT calculations and simulations in this case, e.g., at the position indicated by the arrow in 
Fig.~\ref{fig:Figure13}(a1). However, the peaks due to local orderings are less pronounced if more components are considered. Therefore, Figs.~\ref{fig:Figure13}(b1) 
and \ref{fig:Figure13}(c1) show a much better agreement between simulations and theory. 
This result is confirmed by Figs.~\ref{fig:Figure13}(a2), \ref{fig:Figure13}(b2), and \ref{fig:Figure13}(c2), where 
we compare the density profiles obtained from simulations and DFT. The smoothing of 
these profiles, while increasing the number of components, is the result 
of the increasing number of possible configurations of different stackings next to the wall. 
As a consequence, the peaks are smeared out for an increasing number of components and the splitting 
of a peak can no longer be observed in the case of a more homogeneous spectrum in the polydispersity distribution. 
Nevertheless, Figs.~\ref{fig:Figure13}(b1) and \ref{fig:Figure13}(c1) already show the trend 
that prominent peaks in the pair correlations still occur in the polydisperse situation even for the second shell of surrounding particles. 
These peaks are retained even if the averaged correlation functions 
$h_1(r,z,z')=\tfrac{1}{n}\sum_{\nu=1}^{n} h_{\nu1}^{(2)}(r,z,z')$ (not shown here) would be 
plotted instead of the self-correlations between solely the smallest particles. 
Obviously, these peaks represent the most probable positions of next-neighboring particles, no matter what size they are.

%%%%%%%%%%%%%%%%%%%%%%%%%%%%%%%
%SECTION 6 %%%%%%%%%%%%%%%%%%%%
%%%%%%%%%%%%%%%%%%%%%%%%%%%%%%%
\section{Conclusions}
\label{sec:section6}

Using comparisons to BD simulations, we have quantitatively explored the strengths and weaknesses 
of the WBII FMT approach within DFT in predicting one- and two-particle correlations within HS 
systems. In order to study anisotropic situations, we broke the symmetry and explored the 
behavior in the vicinity of a hard wall. Especially in the case of our six-component systems, DFT led 
to excellent predictions even at high packing fractions. However, in the case of mono- or bidisperse 
systems, DFT did not necessarily resolve the formation of local order. We have demonstrated that the 
compressibility route of DFT can be employed to calculate two-particle correlations, contact values, 
and forces acting on a particle, even in the investigated strongly anisotropic situations.
Our research sets the course for further investigations of structural properties, e.g., within the primitive model, 
where long-ranged particle interactions are involved. Furthermore, it demonstrates the interest in further detailed 
studies on the consistency of functionals. 

Our finding that, particularly at packing fractions above $\phi=0.5$, two-particle correlations can be 
well predicted might turn out to be important to understand the relation of structure and dynamics 
of such systems. For these large packing fractions the dynamics tends to become very slow. Such 
a dramatic slowdown of dynamics usually is termed glassy dynamics and its relation to structure 
is the subject of intensive research \cite{Kirkpatrick1987,Fuchs2002,Schweizer2004,goetze2008complex,Mittal2008,Nygard2012,Mirigian2013,Nygard2014,janssen14,Royall2015a}. 
Advanced theories that deal with glassy dynamics, e.g., mode coupling theory \cite{Fuchs2002,goetze2008complex}, its generalization \cite{janssen14}, 
and similar approaches \cite{Kirkpatrick1987,Schweizer2004,Mirigian2013}, 
rely on the knowledge of the structure of the system. Our work demonstrates that FMT is a suitable 
approach to obtain a reliable input for these theories even in the case of anisotropic geometries, 
e.g., in the vicinity of a wall. 
Furthermore, a comparison of our results to simulations of soft particles away from the HS limit probably is interesting, especially 
for large packing fractions, where the softness of the particle might change the behavior significantly \cite{zhang_nature459_2009,yang_jcp134_2011}. 

In principle, our DFT calculations can be extended to describe the orderings of particles in 
gravity \cite{loewen13,Kohl2014} or of particles possessing charges \cite{haertel_jpcm27_2015}, which might be confined by charged surfaces \cite{Grandner2009}. 
Such extensions lead to systems with many important applications, e.g., 
the formation and in-plane structure of electric double layers \cite{merlet_jpcc118_2014} 
or interfaces like the liquid-vapor one \cite{parry_jpcm26_2014}. 
The knowledge of structural correlations in the so-called (restricted) primitive model 
might lead to advanced insights into the properties of modern devices like 
supercapacitors \cite{simon2008materials,jiang_jpcl3_2012,merlet_jpcc118_2014}, 
blue engines \cite{norman_science186_1974,janssen_prl113_2014,haertel_jpcm27_2015}, and 
thermocapacitive heat-to-current converters \cite{haertel_ees8_2015}.

\section*{Acknowledgements}  
We thank R. van Roij, H. L\"{o}wen, and R. Evans
for stimulating discussions. A.H. acknowledges financial support from the D-ITP consortium, a program
of the Netherlands Organisation for Scientific Research
(NWO) that is funded by the Dutch Ministry of Education, Culture and Science (OCW), from an NWO-VICI grant, and from the Deutsche Forschungsgemeinschaft (DFG) within the priority program
SPP 1726 (Grant No. SP 1382/3-1). M.K. and M.S.
were supported within the Emmy Noether program of the
DFG (Grant No. Schm 2657/2).

%%%%%%%%%%%%%%%%%%%%%%%%%%%%%%%
%SECTION A %%%%%%%%%%%%%%%%%%%%
%%%%%%%%%%%%%%%%%%%%%%%%%%%%%%%
\appendix

\section{Solving the Ornstein-Zernike relation}
\label{sec:appendix1}

Starting with the direct correlation functions $c_{\nu\nu'}^{(2)}$ determined from FMT, 
we obtain the total pair-correlation functions $h_{\nu\nu'}^{(2)}$ by solving the OZ relation 
as defined in Eq.~(\ref{eq:ornstein-zernike-equation}) numerically. 
If the involved correlation functions are rescaled by a factor 
$\sqrt{\rho_{\nu}(\vec{r})\rho_{\nu'}(\vec{r}\,')}$, the result is $0$ in all locations that 
must not be reached by a particle. Therefore, it is sufficient to solve the OZ relation 
only outside of the wall, even if the original direct correlations might be nonzero inside the wall. 

As shown in Eq.~(\ref{eq:ornstein-zernike-equation-cylindrical}), 
we solve the OZ relation numerically in Fourier space, where convolutions become simple products. 
In our case, we consider functions with radial symmetry, i.e., functions $f(x,y)$ 
with $x=r\cos(\theta)$ and $y=r\sin(\theta)$ that do not depend on $\theta$. 
Then their Fourier transforms are 
\begin{equation}
F(f)(k_x,k_y) = \frac{1}{2\pi} \int_{\mathcal{R}^2} f(x,y)
e^{-\imath(x k_x+y k_y)} dx dy,
\end{equation}
which, in polar coordinates after the integration over $\theta$, lead to
\begin{equation}
F(f)\left(s\right) = \int_0^{\infty} r f(r) J_0(s r) dr \,.
\end{equation}
This result corresponds to a Hankel transform (or Bessel transform) as introduced in 
Eq.~(\ref{eq:Hankel-transform}), which in general is defined 
by \cite{johnson_CompPhysCom43_1987,lemoine_jcp101_1994} 
\begin{equation}
F_\nu(u) = H_\nu\left\{f(t)\right\} 
= \int_0^{\infty} f(t) J_\nu(u t) t dt  \label{eq:hankel_transform_definition}
\end{equation}
where $F_\nu(u)$ is called the Hankel transformed function of order $\nu$ of the function $f$ 
if the integral exists. The function $f$ can be a complex-valued function and $J_\nu$ denote 
Bessel functions of the first kind, which, for integer $\nu$, are given 
by \cite{press_book_1992,johnson_CompPhysCom43_1987,lemoine_jcp101_1994} 
\begin{align}
J_\nu(x) &= \frac{1}{2\pi}\int_{-\pi}^\pi e^{-\imath\big(\nu\tau-x\sin(\tau)\big)}d\tau . 
\end{align}
The inverse Hankel transform is given by
\begin{equation}
f(t) = H_\nu^{-1}\left\{F_\nu(u)\right\} 
= \int_0^{\infty} F_\nu(u) J_\nu(u t) u du. 
\label{eq:inverse_hankel_transform_definition}
\end{equation}
We employed the Hankel transform, which, for numerical calculations, 
is available in the {\it Gnu Scientific Library} (GSL) and whose calculation scheme follows 
the work of H.~F.~Johnson \cite{johnson_CompPhysCom43_1987} and 
D.~Lemoine \cite{lemoine_jcp101_1994}.

\section{The weight-correlation functions in FMT}
\label{sec:appendix2}

In this Appendix we derive the terms that are used in FMT for a multicomponent system in order 
to obtain the direct correlation functions. 

From Eq.~(\ref{eq:c2:stepwise-constant}) we know that the direct correlation functions in FMT 
on a discrete numerical grid read 
\begin{equation}
-c_{\nu\nu'}^{(2)}(\vec{r},\vec{r}') 
\approx \sum_{i=0}^{M-1} \sum_\alpha \sum_\beta 
\frac{\partial^2\Phi(z_i)}{\partial n_\alpha\partial n_\beta}
W_{\nu\nu'}^{(\alpha\beta)}(\bar{I}_i, \vec{\Delta}) , 
\label{eq:num-c2}
\end{equation}
where $\vec{\Delta}=\vec{r}'-\vec{r}$,
$z_i$ are the discrete and equidistant sample points along the $z$ axis separated by $d_\text{z}$, 
the weight-correlation functions $W_{\nu\nu'}^{(\alpha\beta)}(I,\vec{r})$ were defined 
in Eq.~(\ref{eq:c2:define-WW}), and $\bar{I}_i$ is a corresponding 
interval, $\bar{I}_i=[z_L,z_R]$, with 
$z_L=z_i-(\vec{r})_z-\tfrac{1}{2}d_\text{z}$ and $z_R=z_i-(\vec{r})_z+\tfrac{1}{2}d_\text{z}$, 
which contains $z_i$. Note that we employ $r_\chi\equiv(\vec{r})_\chi$ as shorthand for 
the $\chi$ component of the vector $\vec{r}$ in Cartesian coordinates spanned by 
$\{\hat{e}_x,\hat{e}_y,\hat{e}_z\}$. 

The weight-correlation functions $W_{\nu\nu'}^{(\alpha\beta)}(I,\vec{r})$ are representations of convolutions of the translational-invariant weight functions $w_\nu^{(\alpha)}$ and 
$w_{\nu'}^{(\beta)}$ from Eqs.~(\ref{eq:weight3})-(\ref{eq:weight2tensor}) on the interval $I$. 
These weight functions have nonvanishing values 
solely on the volume $S_\nu$ or on the surface $\partial S_\nu$ of a sphere of species $\nu$ with radius $R_\nu$.
Thus, we consider two spheres, $A$ and $B$, with centers in the origin and at $\vec{\Delta}$.

\begin{figure*}
\centering
\includegraphics[width=16.0cm]{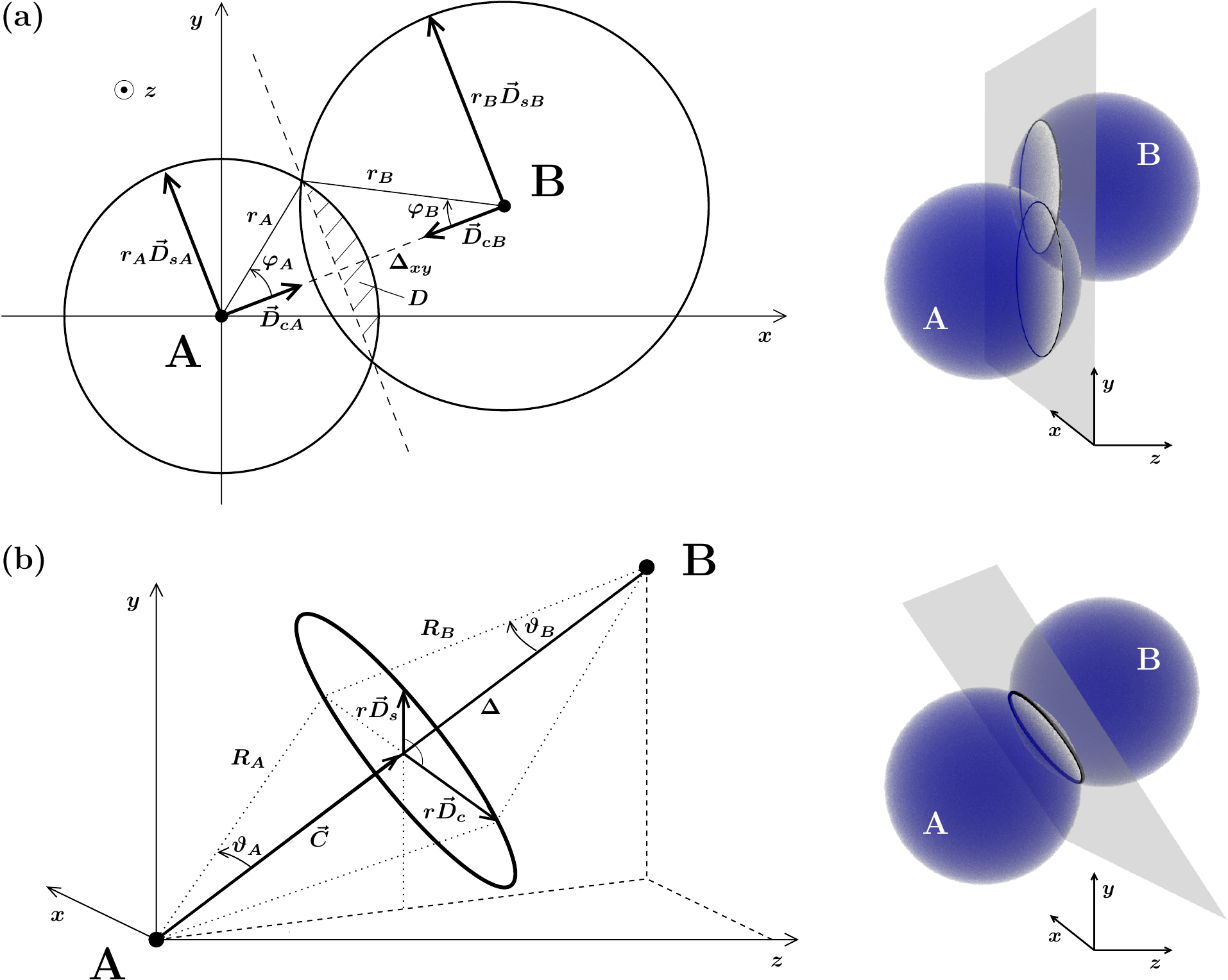}
\caption{\label{fig:c2_prep}(Color online) 
Sketch of the intersection of two spheres $A$ and $B$ with radii $R_A$ and $R_B$ 
at a center-center distance of $\Delta$. 
The sketch contains notations and parametrizations for (a) the intersection in the xy plane and 
(b) the intersection line, both illustrated at the right. 
Note that in (b) $|\vec{C}|=\Delta_A$ and $\Delta=\Delta_A+\Delta_B$. 
}
\end{figure*}

In order to calculate a function $W_{\nu\nu'}^{(\alpha\beta)}$ as given in 
Eq.~(\ref{eq:c2:define-WW}), its integration interval $I$ must have certain properties. 
To guarantee these properties, the interval $I$ can be split into parts 
$I_1$ and $I_2$ with $I_1\cap I_2=\emptyset$ and $I=I_1\cup I_2$, such that 
\begin{equation}
W_{\nu\nu'}^{(\alpha\beta)}(I, \vec{\Delta}) := 
W_{\nu\nu'}^{(\alpha\beta)}(I_1, \vec{\Delta}) + W_{\nu\nu'}^{(\alpha\beta)}(I_2, \vec{\Delta}) . 
\label{eq:splitting}
\end{equation}
Subsequently, splitting $I$ in an appropriate way into intervals $I_i$ guarantees the following 
necessary properties  
after splitting: 
\begin{itemize}
\item Either the weight-correlation function vanishes in the interval $I_i$ 
[$W_{AB}^{(\alpha\beta)}(I_i,\vec{\Delta})=0$] or both spheres, $S_A$ and $S_B$, contain at least 
one point with $z$ component $z$ for each point $z$ in the interval $I_i$ 
($\forall z\in I_i$, $\mathcal{V}_z:=\mathcal{R}^2\times\{z\}$: 
$\mathcal{V}_z\cap S_A\neq\emptyset$ and $\mathcal{V}_z\cap S_B\neq\emptyset$). 
\item Either the intersection $\partial S_A \cap \partial S_B$ of both spheres, $\partial S_A$ and $\partial S_B$, 
contains, for all $z$ in $I_i$, at least one point $\vec{r}$ with $z$ component $r_z$ 
or it contains, for all $z$ in the inner kernel $\mathring{I_i}$, no point $\vec{r}$ with $z$ component $r_z$. 
\end{itemize}
Note that the whole intersection line $\partial S_A\cap\partial S_B$ can be 
contained in one slice, $\mathcal{V}_z:=\mathcal{R}^2\times\{z\}$, 
when $\vec{\Delta} || \hat{e}_z$ (for visualization see Fig.~\ref{fig:c2_prep}). 
We do not consider the special situation where the spheres touch in a single point, which 
would contribute only to the point of the direct correlation function at particle contact, 
whose value is not defined. 

\newcommand{\sign}[0]{\mathrm{sign}}
As can be seen from its definition in Eq.~(\ref{eq:c2:define-WW}), 
the absolute value of the weight-correlation function $W_{AB}^{(\alpha\beta)}$ 
does not change if the spheres $S_A$ and $S_B$ exchange their positions and the interval $I$ 
is adapted in an appropriate way; i.e., $I=[z_L,z_R]$ must be adapted to 
$I'=[(\vec{\Delta})_z-z_R,(\vec{\Delta})_z-z_L]$. However, 
the sign of the function changes when one of the involved weight functions 
is antisymmetric and $\sign(w_A)\sign(w_B)<0$; in our FMT approach, only the 
vectorial weight functions are antisymmetric. Therefore, an exchange of the two spheres leads to 
\begin{equation}
W_{AB}^{(\alpha\beta)}(I,\vec{\Delta}) 
= \sign(w_A^{(\alpha)}) \sign(w_B^{(\beta)}) W_{BA}^{(\beta\alpha)}(I',\vec{\Delta}) . 
\label{eq:weight-correlation-function-symmetry}
\end{equation}
For this reason we calculate only combinations with $\alpha \geq \beta$, 
according to the order $3>2>1>0>\vec{2}>\vec{1}>\tensor{2}$. 
Furthermore, from the definition of the weight functions it follows that 
\begin{align}
W_{AB}^{(\alpha 1)} &= \tfrac{1}{4\pi R_B}W_{AB}^{(\alpha 2)} , 
\label{eq:weight-correlation-function-31} \\
W_{AB}^{(\alpha 0)} &= \tfrac{1}{4\pi R_B^2}W_{AB}^{(\alpha 2)} , \\
W_{AB}^{(\alpha\vec{1})} &= \tfrac{1}{4\pi R_B}W_{AB}^{(\alpha\vec{2})} . 
\label{eq:weight-correlation-function-v1x}
\end{align}
In summary, we have to calculate only the weight-correlation functions for the following combinations:
\begin{align}
(\alpha\beta)\in\big\{&(33),(32),(3\vec{2}),(3\tensor{2}), \notag \\
&(22),(2\vec{2}),(2\tensor{2}),
(\vec{2}\vec{2}),(\vec{2}\tensor{2}),(\tensor{2}\tensor{2})\big\} \,. 
\label{eq:remaining-situations}
\end{align}
All other combinations can be obtained by the relations mentioned above.

If the support of $W_{AB}^{(\alpha\beta)}$ and the volume $\mathcal{V}=\mathcal{R}^2\times I$ do 
overlap (have a nonvanishing intersection), 
three cases are left for this volume $\mathcal{V}$. 
\begin{itemize}
\item[1.] \textbf{Sphere B inside sphere A}\\
      Sphere B is completely encapsulated by sphere A (or vice versa), i.e., without loss of generality,\\
			$S_A\cap S_B\cap\mathcal{V}=S_B\cap\mathcal{V}$ and 
			$\partial S_A\cap S_B\cap\mathring{\mathcal{V}}=\emptyset$. 
\item[2.] \textbf{Partial intersection}\\
      Different spheres with only partial intersection, i.e., without loss of generality,\\
			$\partial S_A\cap\partial S_B\cap\mathring{\mathcal{V}}\neq\emptyset$, but $S_A\neq S_B$. 
\item[3.] \textbf{Two equal spheres}\\
Equally sized spheres are at the same position,\\
      $S_A=S_B$. 
\end{itemize}

In the following sections we calculate the weight-correlation functions 
$W_{AB}^{(\alpha\beta)}$ in this three cases for all combinations of weight functions 
mentioned in Eq.~(\ref{eq:remaining-situations}). During this calculation, we use the in-plane 
radii $r_A$ and $r_B$ of the spheres intersecting with a plane $\mathcal{V}_z$ 
perpendicular to the $z$ axis, i.e., of the circles $\mathcal{V}_z\cap S_A$ and $\mathcal{V}_z\cap S_B$ 
as shown in Fig.~\ref{fig:c2_prep}(a). 
In our three cases, these radii are well defined for all $z\in I$ with planes $\mathcal{V}_z$ within 
the volume $\mathcal{V}=\mathcal{R}^2\times I$ of integration and read 
\begin{align}
r_A(z) &= \sqrt{R_A^2-z^2} \label{eq:inplane-radius-A} , \\
r_B(z) &= \sqrt{R_B^2-\big((\vec{\Delta})_z-z\big)^2} . \label{eq:inplane-radius-B}
\end{align}
Keep in mind that appropriate splitting must guarantee the earlier-mentioned properties after 
Eq.~(\ref{eq:splitting}).

\subsection*{Case 1: Sphere B inside sphere A}
\label{sec:c2:partI}

This case occurs only when A is larger than B and when B is fully encapsulated. 
In this situation, the unit vectors pointing from the centers of sphere $A$ or $B$ 
towards their respective surfaces can be parametrized for $z\in I_i$ 
with cylindrical coordinates $(\gamma,z)$ by 
\begin{align}
\frac{\vec{R}_A(\gamma,z)}{R_A}&=\frac{1}{R_A}
\begin{pmatrix}r_A(z)\cos(\gamma) \\ r_A(z)\sin(\gamma) \\ z\end{pmatrix} , \\
%\end{equation}
%\begin{equation}
\frac{\vec{R}_B(\gamma,z)}{R_B}&=\frac{1}{R_B}
\begin{pmatrix}-r_B(z)\cos(\gamma) \\ r_B(z)\sin(\gamma) \\ z-(\vec{\Delta})_z\end{pmatrix} ,  
\end{align}
where $r_A(z)$ and $r_B(z)$ are given by Eqs. (\ref{eq:inplane-radius-A}) and (\ref{eq:inplane-radius-B}).

For all combinations, where the weight function of the larger encapsulating sphere is not 
$w_A^{(3)}$, neither weight function intersects and one trivially obtains 
$W_{AB}^{(\alpha\neq3,\beta)}=0$; we neglect the case where the encapsulated sphere touches the outer 
one in a single point. For the remaining combinations of weight functions the first two weight-correlation 
functions read 
\begin{align}
W_{AB}^{(33)} &= \int_{z_L}^{z_R} \pi r_B^2(z)dz \notag \\
&= \left[ \pi R_B^2 z + \frac{\pi}{3}\Big((\vec{\Delta})_z-z\Big)^3 \right]_{z=z_L}^{z_R} , \\
W_{AB}^{(32)} &= \int_{\mathcal{V}_i} 
\Theta\big(R_A-|\vec{r}|\big) \delta\big(R_B-|\vec{r}-\vec{\Delta}|\big) d\vec{r} . \label{eq:C1-W32-1}
\end{align}
Since sphere $B$ is encapsulated inside of sphere $A$, the $\Theta$ weight in 
Eq.~(\ref{eq:C1-W32-1}) is equal to unity for the integration volume of interest. Furthermore, 
a linear parameter change for the $xy$ integration in this equation and a change to 
cylindrical coordinates $(r\cos(\gamma),r\sin(\gamma),z)$ lead to 
\begin{equation}
W_{AB}^{(32)} = \int_{I_i} \int_{0}^{2\pi} \int_{0}^{\infty} r 
\delta\Big(R_B-\sqrt{r^2+\big(z-(\vec{\Delta})_z\big)^2}\Big) dr d\gamma dz . \label{eq:C1-W32-2}
\end{equation}
In order to perform the integrals in Eq.~(\ref{eq:C1-W32-2}), we use the equality 
\begin{equation}
\delta\big(g(r)\big) = \sum_{i}\frac{\delta(r-r_i)}{|g'(r_i)|} 
\label{eq:delta-arg-change}
\end{equation}
for a continuously differentiable function $g(r)$ with the finite set $\{r_i\}$ of simple $0$'s 
and the derivative 
$g'(r)=\partial g/\partial r$. In Eq.~(\ref{eq:C1-W32-2}) the argument of the 
$\delta$ distribution has the simple zero $r_1=r_B(z)$ and $|g'(r_1)|=r/R_B$. 
Accordingly, the previous result of Eq.~(\ref{eq:C1-W32-2}) becomes 
\begin{equation}
W_{AB}^{(32)} = 
2\pi R_B (z_R-z_L) . \label{eq:C1-W32-3}
\end{equation}
Similarly, it follows that 
\begin{align}
	W_{AB}^{(3\vec{2})} =& \int_{z_L}^{z_R} \int_{0}^{2\pi} \int_{0}^{\infty} 
R_B\delta\big(r-r_B(z)\big) 
\vec{R}_B(\gamma,z) \, dr d\gamma dz \notag \\
=& 
 2 \pi \hat{e}_z 
 \left[\frac{1}{2}z^2-(\vec{\Delta})_z z\right]_{z=z_L}^{z_R} , \\
W_{AB}^{(3\tensor{2})} 
=& 	\big( \hat{e}_{\rm x}\otimes\hat{e}_{\rm x} + \hat{e}_y\otimes\hat{e}_y \big) \notag \\
&  \quad \times \frac{\pi}{R_B} 
\left[R_B^2 z + \frac{1}{3}\big((\vec{\Delta})_z-z\big)^3\right]_{z=z_L}^{z_R} \notag \\
&  + 
\big(\hat{e}_z\otimes\hat{e}_z\big) \frac{2\pi}{R_B} 
\left[-\frac{1}{3}\big((\vec{\Delta})_z-z\big)^3\right]_{z=z_L}^{z_R} \notag \\
&  - \frac{\tensor{{\rm I}}}{3}W_{AB}^{(32)} , 
\end{align}
where the outer product $\hat{e}_i\otimes\hat{e}_j$ between $\hat{e}_i$ and $\hat{e}_j$ is defined 
as the matrix product $\hat{e}_i\cdot\hat{e}_j^{T}$ with $T$ indicating a transposed vector.

\subsection*{Case 2: Partial intersection}

In this case, both sphere $A$ and sphere $B$ intersect each other and the intersection occurs at $z$ positions with $z_L\leq z\leq z_R$.
In order to calculate the weight-correlation functions $W_{AB}^{(\alpha\beta)}$ we distinguish two 
cases. 
\begin{itemize}
\item[(2a)] At least one of the corresponding weight functions %$w_A^{(\alpha)}$ 
incorporates a $\Theta$ weight: 
$\Leftrightarrow \alpha = 3 \geq \beta$. 
\item[(2b)] No $\Theta$-weight function is involved: 
$\Leftrightarrow 3 > \alpha \geq \beta$. 
\end{itemize}

\subsubsection*{Case 2a: Partial intersection, $\alpha=3$}

In this case, we employed numerical integration in order to determine $W_{AB}^{(\alpha\beta)}$ following some analytical calculations.  

According to previous discussions, 
$\Delta_z:=(\vec{\Delta})_z<|\vec{\Delta}|$ and 
$\Delta_{xy}:=\sqrt{(\vec{\Delta})_x^2+(\vec{\Delta})_y^2}>0$. 
Thus, the vectors $\vec{R}_A$ and $\vec{R}_B$, which point from the center of the spheres 
$S_A$ and $S_B$ to their surface (at position $z$), can be parameterized by 
(see Fig.~\ref{fig:c2_prep}) 
\begin{align}
\vec{R}_A(\varphi_A) &= \vec{C}_A 
 + r_A(z)\left( \vec{D}_{cA}\cos(\varphi_A) + \vec{D}_{sA}\sin(\varphi_A) \right) , \\ 
\vec{R}_B(\varphi_B) &= \vec{C}_B 
 + r_B(z)\left( \vec{D}_{cB}\cos(\varphi_B) + \vec{D}_{sB}\sin(\varphi_B) \right) , 
 \label{eq:C2a-RB}
\end{align}
where
\begin{align}
\vec{C}_A &= z \hat{e}_z 
\textrm{ , }
\vec{C}_B = (z-\Delta_z) \hat{e}_z , \\ 
\vec{D}_{cA} &= \Delta_{xy}^{-1} (\Delta_x\hat{e}_x+\Delta_y\hat{e}_y) 
= -\vec{D}_{cB} , \\
\vec{D}_{sA} &= \hat{e}_z \times \vec{D}_{cA} 
= -\hat{e}_z \times \vec{D}_{cB} = \vec{D}_{sB}.
\end{align}
The in-plane radii $r_A$ and $r_B$ are used as defined in Eqs.~(\ref{eq:inplane-radius-A}) and (\ref{eq:inplane-radius-B}). 
From the law of cosines it follows that 
\begin{align}
r_A \cos(\varphi_A) &= \frac{r_A^2+\Delta_{xy}^2-r_B^2}{2\Delta_{xy}}\textrm{ , }\varphi_A\in(0,\pi) \\
r_B \cos(\varphi_B) &= \frac{r_B^2+\Delta_{xy}^2-r_A^2}{2\Delta_{xy}}\textrm{ , }\varphi_B\in(0,\pi) , 
\end{align}
where the correlated angles $\varphi_A$ and $\varphi_B$ become 
$\tfrac{\pi}{2}$ for vanishing radii $r_A$ and $r_B$, respectively.

In the case of two $\Theta$ weights, the intersection area of the kernel is given 
by two caps of the corresponding intersecting circles as
illustrated in Fig.~\ref{fig:c2_prep}(a). 
The area $D$ of such a cap is given by the fraction $\tfrac{2\varphi}{2\pi}$ of the 
corresponding circle with a triangle subtracted or added, depending on the opening angle of $\varphi$: 
if $\varphi\leq\tfrac{\pi}{2}$, the triangle 
is subtracted; otherwise, it is added. With $h=|\sin(\varphi)|r$, the area follows with 
\begin{equation}
D = \varphi r^2 - r^2\cos(\varphi)\sin(\varphi) . 
\end{equation}
Thus, the weight-correlation function for two $\Theta$ weights follows with 
\begin{align}
W_{AB}^{(33)} = \int_{z_L}^{z_R} \Big( 
& r_A^2(z) \big( \varphi_A - \sin(\varphi_A)\cos(\varphi_A) \big) \label{eq:C2-W33} \\
+ & r_B^2(z) \big( \varphi_B - \sin(\varphi_B)\cos(\varphi_B) \big) \notag
\Big) dz . 
\end{align}

Referring to calculations from Eqs.~(\ref{eq:C1-W32-2}) - (\ref{eq:C1-W32-3}) 
in case 1, we, furthermore, get 
\begin{equation}
W_{AB}^{(32)} = \int_{z_L}^{z_R} \int_{-\varphi_B}^{\varphi_B} R_B d\gamma dz
= 2 \int_{z_L}^{z_R} \varphi_B R_B dz . 
\end{equation}
Using the parametrization of $\vec{R}_B$ from Eq.~(\ref{eq:C2a-RB}), we obtain
\begin{align}
W_{AB}^{(3\vec{2})} &= \int_{z_L}^{z_R} \int_{-\varphi_B}^{\varphi_B} 
\vec{R}_B(\varphi) d\varphi dz \notag \\
&= 2 \int_{z_L}^{z_R} \left[ \varphi_B\vec{C}_B + r_B(z)\vec{D}_{cB}\sin(\varphi_B) \right] dz . 
\end{align}
Using, furthermore, the equalities $\int\sin^2(x)dx=\tfrac{x}{2}-\tfrac{1}{4}\sin(2x)$, 
$\int\cos^2(x)dx=\tfrac{x}{2}+\tfrac{1}{4}\sin(2x)$, 
and $\int\sin(x)\cos(x)dx=-\tfrac{1}{2}\cos^2(x)$, it follows that 
\begin{widetext}
\begin{align}
\left(W_{AB}^{(3\tensor{2})}\right)_{ij} 
	=& \int_{z_L}^{z_R} \int_{-\varphi_B}^{\varphi_B} 
 \frac{R_B}{R_B^2} \left(\vec{R}_B(\varphi)\right)_i\left(\vec{R}_B(\varphi)\right)_j d\varphi dz 
 - \int_{z_L}^{z_R} \int_{-\varphi_B}^{\varphi_B} R_B \frac{1}{3}\delta_{ij} d\varphi dz \\
 =& \int_{z_L}^{z_R} \frac{2}{R_B} \bigg[
\varphi_B \left(\vec{C}_B\right)_i \left(\vec{C}_B\right)_j \notag \\
& +r_B(z)\sin(\varphi_B) \left(\vec{C}_{B}\right)_i \left(\vec{D}_{cB}\right)_j 
 +r_B(z)\sin(\varphi_B) \left(\vec{D}_{cB}\right)_i \left(\vec{C}_B\right)_j \notag \\
& +\big(r_B(z)\big)^2\left(\frac{\varphi_B}{2}+\frac{1}{4}\sin(2\varphi_B)\right) 
\left(\vec{D}_{cB}\right)_i \left(\vec{D}_{cB}\right)_j 
 +\big(r_B(z)\big)^2\left(\frac{\varphi_B}{2}-\frac{1}{4}\sin(2\varphi_B)\right) 
\left(\vec{D}_{sB}\right)_i \left(\vec{D}_{sB}\right)_j \bigg] dz \notag \\
& -\frac{1}{3}\delta_{ij} W_{AB}^{(32)} \, . \label{eq:C2-W3t2-2}
\end{align}
%\begin{align}
%\left(W_{AB}^{(3\tensor{2})}\right)_{ij} 
%	=& \int_{z_L}^{z_R} \int_{-\varphi_B}^{\varphi_B} 
% \frac{R_B}{R_B^2} \left(\vec{R}_B(\varphi)\right)_i\left(\vec{R}_B(\varphi)\right)_j d\varphi dz \notag \\
%  & - \int_{z_L}^{z_R} \int_{-\varphi_B}^{\varphi_B} R_B \frac{1}{3}\delta_{ij} d\varphi dz \\
% =& \int_{z_L}^{z_R} \frac{2}{R_B} \bigg[
%\varphi_B \left(\vec{C}_B\right)_i \left(\vec{C}_B\right)_j \notag \\
%& +r_B(z)\sin(\varphi_B) \left(\vec{C}_{B}\right)_i \left(\vec{D}_{cB}\right)_j \notag \\
%& +r_B(z)\sin(\varphi_B) \left(\vec{D}_{cB}\right)_i \left(\vec{C}_B\right)_j \notag \\
%& +\big(r_B(z)\big)^2\left(\frac{\varphi_B}{2}+\frac{1}{4}\sin(2\varphi_B)\right) 
%\left(\vec{D}_{cB}\right)_i \left(\vec{D}_{cB}\right)_j \notag \\
%& +\big(r_B(z)\big)^2\left(\frac{\varphi_B}{2}-\frac{1}{4}\sin(2\varphi_B)\right) 
%\left(\vec{D}_{sB}\right)_i \left(\vec{D}_{sB}\right)_j \bigg] dz \notag \\
%& -\frac{1}{3}\delta_{ij} W_{AB}^{(32)} \, . \label{eq:C2-W3t2-2}
%\end{align}
\end{widetext}

Finally, we calculated the remaining integral over the interval $I_i=[z_L,z_R]$ in 
Eqs.~(\ref{eq:C2-W33})-(\ref{eq:C2-W3t2-2}) numerically on a discrete grid of 16 points. 
Keep in mind that $z_R-z_L\leq d_\text{z}$, which is the numeric resolution chosen for the 
determination of the direct correlation function in Eq.~(\ref{eq:c2:stepwise-constant}).

\subsubsection*{Case 2b: Partial intersection, $\alpha<3$}

In the interval $I_i=[z_L,z_R]$ of interest, a unique intersection 
circle between the surfaces $\partial S_A$ and $\partial S_B$ exists. 
Note that the whole intersection circle might lie in one plane, $\mathcal{R}^2\times\{z_c\}$, 
if $\vec{\Delta}\parallel\hat{e}_z$. Otherwise, the distance $\vec{\Delta}$ must have nonvanishing 
contributions orthogonal to $\hat{e}_z$.

The intersection circle, as sketched in Fig.~\ref{fig:c2_prep}(b), 
can be parameterized by the vector 
\begin{equation}
\vec{r}_{\rm I}(t) = \vec{C} + \vec{D}_{c}r_{\rm I}\cos(t) + \vec{D}_{s}r_{\rm I}\sin(t) , 
\end{equation}
where the radius $r_{\rm I} = \sqrt{R_A^2-\Delta_A^2}=\sin(\vartheta_A)R_A$ follows 
from $R_B^2=R_A^2+\Delta^2-2R_A\Delta\cos(\vartheta_A)$ with $\Delta\equiv|\vec{\Delta}|$ and 
from $\Delta_A=\cos(\vartheta_A)R_A$.  

For $\vec{\Delta}\nparallel\hat{e}_z$, the vectors in the parametrization read 
\begin{align}
\vec{C} &= \Delta_A\frac{\vec{\Delta}}{|\vec{\Delta}|} 
 = \frac{R_B^2-R_A^2-\Delta^2}{-2\Delta^2} \begin{pmatrix}\Delta_x\\ \Delta_y\\ \Delta_z\end{pmatrix} , \\
\vec{D}_s &= \frac{\hat{e}_z \times\vec{\Delta}}{|\hat{e}_z \times\vec{\Delta}|}
 = \frac{1}{\Delta_{xy}} \begin{pmatrix}-\Delta_y\\ \Delta_x\\ 0\end{pmatrix} , \\
\vec{D}_c &= \frac{\vec{D}_s \times\vec{\Delta}}{|\vec{\Delta}|}
	 = \frac{1}{\Delta_{xy} \Delta} 
 \begin{pmatrix}-\Delta_x\Delta_z\\ \Delta_y\Delta_z\\ -\Delta_{xy}^2\end{pmatrix} .
\end{align}
Moreover, 
$|\vec{D}_s\times\vec{\Delta}|=|\vec{\Delta}|$, because $\vec{D}_s\perp\vec{\Delta}$ and 
$|\vec{D}_s|=1$. By definition, it also follows that $\vec{D}_s\perp\vec{D}_c$. 
To map the parameter $t$ into the given interval $I_i$ we, furthermore, solve 
$z=(\vec{r}_{\rm I}(t))_z$ and find 
\begin{equation}
cos(t) = 
\frac{z-\frac{R_B^2-R_A^2-\Delta^2}{-2\Delta^2}\Delta_z}{-2\Delta_{xy}}\Delta . 
\end{equation}
Thus, the interval $I_i=[z_L,z_R]$ corresponds to the intervals $[t_1,t_2]$ and $[-t_2,-t_1]$, 
due to the symmetry properties of the cosine. 

In the case where $\vec{\Delta}\parallel\hat{e}_z$, when the whole intersection circle is located in 
one $z$ slice at $z=z_c$, we set the vectors in the parametrization to 
$\vec{C}=z_c\hat{e}_z$, $\vec{D}_s=\hat{e}_y$, and $\vec{D}_c=\hat{e}_x$. Then the whole 
circle is caught by the above-defined intervals $[t_1,t_2]$ and $[-t_2,-t_1]$ with 
$t_1=0$ and $t_2=\pi$. 

Now, we consider the weight-correlation function, 
\begin{equation}
W_{AB}^{(22)} = \int_{z_L}^{z_R} \int \int 
\delta\big(R_A-|\vec{r}|\big) \delta\big(R_B-|\vec{r}-\vec{\Delta}|\big) d\vec{r} . 
\label{eq:C2b-W22-1}
\end{equation}
Splitting the vector $\vec{r}$ into parallel and orthogonal components 
$\vec{r}_{\parallel}\parallel\vec{\Delta}$ and $\vec{r}_{\perp}\perp\vec{\Delta}$ and 
converting to cylindrical coordinates $(r\cos(\gamma),r\sin(\gamma),c\equiv|\vec{C}|)$ on the 
Euclidean base $(\vec{D}_c,\vec{D}_s,\vec{C}/c)$, we find 
\begin{align}
W_{AB}^{(22)} =& \int_{\mathcal{R}} \int_{-t_2}^{-t_1} \int_{0}^{\infty} 
r \delta\big(g_A(c)\big) \delta\big(g_B(r)\big) dr d\gamma dc \notag \\
+& \int_{\mathcal{R}} \int_{t_1}^{t_2} \int_{0}^{\infty} 
r \delta\big(g_A(c)\big) \delta\big(g_B(r)\big) dr d\gamma dc ,
\label{eq:C2b-W22-2}
\end{align}
with the argument functions $g_A(c)=R_A-\sqrt{r^2+c^2}$ and 
$g_B(r)=R_B-\sqrt{r^2+\big(c-|\vec{\Delta}|\big)^2}$, 
where the conditions concerning the $z$ integration from Eq.~(\ref{eq:C2b-W22-1}) have been transferred 
to conditions of the $\gamma$ integration. 

In this tilted geometry,  
we first apply the identity 
from Eq.~(\ref{eq:delta-arg-change}) to the second $\delta$ distribution with argument $g_B(r)$ 
in Eq.~(\ref{eq:C2b-W22-2}) and 
achieve the simple zero $r_0=\sqrt{R_B^2-\big(c-|\vec{\Delta}|\big)^2}$ together with 
$|g'_B(c_0)|=r/R_B$. 
Second, we apply the same identity to the first $\delta$ distribution with argument $g_A(c)$, 
where we already replaced the parameter $r$ with the value which is set by the $r$ integration over 
the second $\delta$ distribution, leading to $g_A(c)=R_A-\sqrt{c^2+R_B^2-\big(c-|\vec{\Delta}|\big)^2}$, 
with the simple zero $c_0=\big(R_A^2-R_B^2+|\vec{\Delta}|^2\big)/(2|\vec{\Delta}|)$ and the 
corresponding $|g'_A(c_0))|=|\vec{\Delta}|/R_A$. Accordingly, we find 
\begin{align}
W_{AB}^{(22)} &= \int_{\mathcal{R}} \int_{-t_2}^{-t_1} \int_{0}^{\infty} \frac{R_A R_B}{\Delta} 
\delta(c-c_0) \delta(r-r_{\rm I}) dr d\gamma dc \notag \\
&+ \int_{\mathcal{R}} \int_{t_1}^{t_2} \int_{0}^{\infty} \frac{R_A R_B}{\Delta} 
\delta(c-c_0) \delta(r-r_{\rm I}) dr d\gamma dc , 
\label{eq:C2b-W22-4}
\end{align}
which leads to the final result, 
\begin{equation}
W_{AB}^{(22)} = \frac{R_A R_B}{|\vec{\Delta}|} 2 (t_2-t_1) . 
\end{equation}

The vectorial and tensorial weight-correlation functions are calculated in a similar manner. For this purpose, 
we define vectors 
\begin{equation}
\vec{R}_A(t)=\vec{r}_{\rm I}(t)
\textrm{ and }
\vec{R}_B(t)=\vec{r}_{\rm I}(t)-\vec{\Delta} 
\end{equation}
which point from the centers of the spheres $A$ and $B$ to a point on the intersection line 
$\partial S_A \cap\partial S_B$, which is parameterized by t. 
In combination with Eq.~(\ref{eq:C2b-W22-4}) we obtain 
\begin{align}
W_{AB}^{(2\vec{2})} 
&= \frac{R_A R_B}{|\vec{\Delta}|} \int_{-t_2}^{-t_1} 
\frac{\vec{R}_B(\gamma)}{R_B} d\gamma 
  + \frac{R_A R_B}{|\vec{\Delta}|} \int_{t_1}^{t_2} 
\frac{\vec{R}_B(\gamma)}{R_B} d\gamma \notag \\
&= 2\frac{R_A}{|\vec{\Delta}|} \left[ 
 \left(\vec{C}-\vec{\Delta}\right) t + r_0\vec{D}_c \sin(t) \right]_{t=t_1}^{t_2} , 
\end{align}
\begin{align}
\left(W_{AB}^{(2\tensor{2})}\right)_{ij} 
&= \frac{R_A R_B}{|\vec{\Delta}|} \int_{-t_2}^{-t_1} 
 \left(\tfrac{\left(\vec{R}_B(\gamma)\right)_i}{R_B}\tfrac{\left(\vec{R}_B(\gamma)\right)_j}{R_B} 
 -\frac{\delta_{ij}}{3}\right) d\gamma \notag \\
&+ \frac{R_A R_B}{|\vec{\Delta}|} \int_{t_1}^{t_2} 
 \left(\tfrac{\left(\vec{R}_B(\gamma)\right)_i}{R_B}\tfrac{\left(\vec{R}_B(\gamma)\right)_j}{R_B} 
 -\frac{\delta_{ij}}{3}\right) d\gamma , \label{eq:C2b-W22t} \\
\left(W_{AB}^{(\vec{2}\vec{2})}\right)_{ij} 
&= \frac{R_A R_B}{|\vec{\Delta}|} \int_{-t_2}^{-t_1} 
 \tfrac{\left(\vec{R}_A(\gamma)\right)_i}{R_A}
 \tfrac{\left(\vec{R}_B(\gamma)\right)_j}{R_B} d\gamma \notag \\
&+ \frac{R_A R_B}{|\vec{\Delta}|} \int_{t_1}^{t_2} 
 \tfrac{\left(\vec{R}_A(\gamma)\right)_i}{R_A}
 \tfrac{\left(\vec{R}_B(\gamma)\right)_j}{R_B} d\gamma , 
\end{align}
\begin{align}
& \left(W_{AB}^{(\vec{2}\tensor{2})}\right)_{ijk} \notag \\
& = \frac{R_A R_B}{|\vec{\Delta}|} \int_{-t_2}^{-t_1} 
 \tfrac{\left(\vec{R}_A(\gamma)\right)_i}{R_A} 
 \left(
 \tfrac{\left(\vec{R}_B(\gamma)\right)_j}{R_B} 
 \tfrac{\left(\vec{R}_B(\gamma)\right)_k}{R_B} 
 -\frac{\delta_{jk}}{3} \right) d\gamma \notag \\
& + \frac{R_A R_B}{|\vec{\Delta}|} \int_{t_1}^{t_2} 
 \tfrac{\left(\vec{R}_A(\gamma)\right)_i}{R_A} 
 \left(
 \tfrac{\left(\vec{R}_B(\gamma)\right)_j}{R_B} 
 \tfrac{\left(\vec{R}_B(\gamma)\right)_k}{R_B} 
 -\frac{\delta_{jk}}{3} \right) d\gamma , 
\end{align}
\begin{align}
& \left(W_{AB}^{(\tensor{2}\tensor{2})}\right)_{ijkl} \notag \\
& = \frac{R_A R_B}{|\vec{\Delta}|} \int_{-t_2}^{-t_1} 
 \left(\tfrac{\left(\tensor{R}_A(\gamma)\right)_{ij}}{R_A^2} 
 -\frac{\delta_{ij}}{3} \right) 
 \left(\tfrac{\left(\tensor{R}_B(\gamma)\right)_{kl}}{R_B^2}
 -\frac{\delta_{kl}}{3} \right) d\gamma \notag \\
& + \frac{R_A R_B}{|\vec{\Delta}|} \int_{t_1}^{t_2} 
 \left(\tfrac{\left(\tensor{R}_A(\gamma)\right)_{ij}}{R_A^2} 
 -\frac{\delta_{ij}}{3} \right) 
 \left(\tfrac{\left(\tensor{R}_B(\gamma)\right)_{kl}}{R_B^2}
 -\frac{\delta_{kl}}{3} \right) d\gamma \label{eq:C2b-W2t2t} , 
\end{align}
where 
$\tensor{R}_A(\gamma) = \vec{R}_A(\gamma) \otimes \vec{R}_A(\gamma)$ 
with the tensor product $\otimes$, 
$\left(\tensor{R}_A(\gamma)\right)_{ij} 
= \left(\vec{R}_A(\gamma)\right)_{i} \left(\vec{R}_A(\gamma)\right)_{j}$, 
and 
$\left(\tensor{R}_B(\gamma)\right)_{ij} 
= \left(\vec{R}_B(\gamma)\right)_{i} \left(\vec{R}_B(\gamma)\right)_{j}$. 
The analytical form of Eqs.~(\ref{eq:C2b-W22t})-(\ref{eq:C2b-W2t2t}) follows 
from straight forward integration.

\subsection*{Case 3: Two equal spheres}

In the last case, sphere $B$ is equal to sphere $A$.\label{sec:app-case1} This case of equal 
spheres corresponds to a limiting case of the first two cases such that we do not need additional 
calculations. For example, case 1 already covers all situations where $\alpha=3$. These 
situations are addressed, when, in the discussion 
in Sec.~\ref{sec:text-case1}, the correlations between a small and a large particle are 
called similar to the self-correlations of the small particles. In this discussion all cases 
with $\alpha\leq 2$ were neglected. In cases with $\alpha=2$, we find 
\begin{equation}
W_{AB}^{(22)} = \int_{z_L}^{z_R} \int \int 
\delta\left(R_A-\sqrt{r^2+z^2}\right) \delta(R_B-R_A) d\vec{r} .
\end{equation}
This result corresponds to Eq.~(\ref{eq:C1-W32-1}) in case 1, where $\alpha=3$ and $\beta=2$, because 
the $\Theta$-weight of sphere $A$ completely contains the weight function of sphere $B$ 
and, as a consequence, is irrelevant. Note that here the naming of spheres $A$ and $B$ was switched. 

All remaining situations with $\alpha<2$ can be mapped onto the situation where $\alpha=2$, 
because all weights with $\alpha<3$ are $\delta$ weights and differ only in a prefactor. This 
applies even for the vectorial and tensorial weights: 
for example, when $\alpha=\beta=\vec{2}$, both vectors always point to the same point on the 
surface of both spheres such that they are parallel. Accordingly, the result is equal to 
the result obtained for $\alpha=\beta=2$. Similarly, a vectorial and a tensorial weight 
can be reduced to a scalar and a vectorial one, two tensorial ones can be reduced to two 
scalar ones, etc. In conclusions, all combinations of $\delta$ weights can be mapped 
onto the $\alpha=2$ situation.

%\bibliography{references,literature,literature_new,references_matthias_extra,literature_recent}

%merlin.mbs apsrev4-1.bst 2010-07-25 4.21a (PWD, AO, DPC) hacked
%Control: key (0)
%Control: author (8) initials jnrlst
%Control: editor formatted (1) identically to author
%Control: production of article title (-1) disabled
%Control: page (0) single
%Control: year (1) truncated
%Control: production of eprint (-1) disabled
%

\end{document}